# Efficient Route Tracing from a Single Source

Benoit Donnet, Philippe Raoult, Timur Friedman

Université Pierre & Marie Curie, Laboratoire LIP6–CNRS, UMR 7606, Paris, France

## Abstract

Traceroute is a networking tool that allows one to discover the path that packets take from a source machine, through the network, to a destination machine. It is widely used as an engineering tool, and also as a scientific tool, such as for discovery of the network topology at the IP level. In prior work, authors on this technical report have shown how to improve the efficiency of route tracing from multiple cooperating monitors. However, it is not unusual for a route tracing monitor to operate in isolation. Somewhat different strategies are required for this case, and this report is the first systematic study of those requirements. Standard traceroute is inefficient when used repeatedly towards multiple destinations, as it repeatedly probes the same interfaces close to the source. Others have recognized this inefficiency and have proposed tracing backwards from the destinations and stopping probing upon encounter with a previously-seen interface. One of this technical report's contributions is to quantify for the first time the efficiency of this approach. Another contribution is to describe the effect of non-responding destinations on this efficiency. Since a large portion of destination machines do not reply to probe packets, backwards probing from the destination is often infeasible. We propose an algorithm to tackle non-responding destinations, and we find that our algorithm can strongly decrease probing redundancy at the cost of a small reduction in node and link discovery.

## 1 Introduction

*Traceroute* [1] is a networking diagnostic tool natively available on most of the operating systems. It allows one to determine the path followed by a packet. Traceroute allows therefore to draw up the map of router interfaces present along the path between a machine S (the *source* or the *monitor*) and a machine D (the *destination*). Traceroute has also engineering applications as it can be used, for instance, to detect routers that fail in a network. This report proposes and evaluates improvements to standard traceroute for tracing routes from a single point.

Today's most extensive tracing system at the IP interface level, *skitter* [2], uses 24 monitors, each targeting on the order of one million destinations. In the fashion of skitter, *scamper* [3] makes use of several monitors to traceroute IPv6 networks. The *Distributed Internet MEasurements & Simulations* [4] (DIMES) is a measurement infrastructure somewhat similar to the famous SETI@home [5]. SETI@home's screensaver downloads and analyzes radio-telescope data. The idea behind DIMES is to provide to the research community a publicly downloadable distributed route tracing tool. It was released as a daemon in September 2004. The DIMES agent performs Internet measurements such as traceroute and ping at a low rate, consuming at peak 1KB/sec. At the time of writing this report, DIMES counts more than 8,700 agents scattered over five continents. In the fashion of skitter, *scamper* [3] makes use of several monitors to traceroute IPv6 networks. Other well known systems, such as RIPE *NCC TTM* [6] and NLANR *AMP* [7], each employs a larger set of monitors, on the order of one- to two-hundred, but they avoid probing outside their own network. *Scriptroute* [8] is a system that allows an ordinary internet user to perform network measurements from several distributed vantage points. It proposes remote measurement execution on PlanetLab nodes [9], through a daemon that implements ping, hop-by-hop bandwidth measurement, and a number of other utilities in addition to traceroute.

Recently, in the context of large-scale internet topology discovery, we have shown [10] that standard traceroute probing (such as skitter) is particularly inefficient due to duplication of effort at two levels: measurements made by an individual monitor that replicate its own work (*intra-monitor* redundancy), and measurements made by multiple monitors that replicate each other's work (*inter-monitor* redundancy). Using skitter data from August 2004, we have quantified both kinds of redundancy. We showed that intra-monitor redundancy is high close to each monitor and, with respect to inter-monitor redundancy, we find that most interfaces are visited by all monitors, especially when close to destinations. We further proposed an algorithm, *Doubletree*, for reducing both forms of redundancy at the same time.

This technical report focuses more deeply on the intra-monitor redundancy problem. Systems that discover internet topology at IP level from a set of isolated vantage points (i.e., there is no cooperation between monitors) have interest to reduce their intra-monitor redundancy. By sending much less probes, monitors can probe the network more frequently. The more frequent snapshots you have, the more accurate should be your view of the topology. This technical report demonstrates how a monitor can act to reduce its intra-monitor redundancy.

The nature of intra-monitor redundancy suggest to start probing far from the traceroute monitor and probe backwards (i.e., decreasing TTLs), as first noticed by Govindan and Tangmunarunkit [11], Spring et al. [8], Moors [12] and Donnet et al. [10]. However, performing backward probing from non-cooperative traceroute monitors in the context of intra-monitor redundancy has never been evaluated previously. Even if backward probing is simple to understand, it is not clear how efficient it is. This report evaluates the redundancy reduction of backward probing as well as the eventual information lost compared to standard traceroute.

Nevertheless, backward probing is based on the assumption that destinations reply to probes in order to estimate path lengths and the distance of the last hop before the destination. Unfortunately, a large set of destinations (40% in our data set) does not reply to probes, probably due to strongly configured firewalls. In this case, backward probing cannot be performed. In this report, we also propose a way to face non-responding destinations. We further propose an efficient algorithm that can handle both cases, i.e., responding and non-responding destinations. We evaluate these algorithms in terms of intra-monitor redundancy and quantity of information lost.

The remainder of this report is organized as follows: Sec. 2 introduces the data set used throughout this technical report; Sec. 3 gives a key for reading quantile plots; Sec. 4 evaluates standard traceroute; Sec. 5 presents and evaluates separately our backward probing algorithms; Sec. 6 compares the different algorithms; finally, Sec. 7 concludes this report by summarizing its principal contributions.

## 2 Data Set

Our study is based on skitter [2] data from August $1^{st}$ through $3^{rd}$, 2004. This data set was generated by 24 monitors located in the United States, Canada, the United Kingdom, France, Sweden, the Netherlands, Japan, and New Zealand. The monitors share a common destination set of 971,080 IPv4 addresses. Each monitor cycles through the destination set at its own rate, taking typically three days to complete a cycle. For the purpose of our studies, in order to reduce computing time to a manageable level, we worked from a limited destination set of 50,000, randomly chosen from the original set.

Visits to host and router interfaces are the metric by which we evaluate redundancy. We consider an interface to have been visited if its IP address returned by the host or router appears, at least, at one of the hops in a traceroute. Though it would be of interest to calculate the load at the host and router level, rather than at the individual interface level, we make no attempt to disambiguate interfaces in order to obtain router-level information. The alias resolution techniques described by Pansiot and Grad [13], by Govindan and Tangmunarunkit [11], for *Mercator*, and applied in the *iffinder* tool from CAIDA [14], would require active probing beyond the skitter data, preferably at the same time that the skitter data is collected. The methods used by Spring et al. [15], in *Rocketfuel*, and by Teixeira et al. [16], apply to routers in the network core, and are untested in stub networks. Despite these limitations, we believe that the load on individual interfaces is a useful measure. As Broido and claffy note [17], "interfaces are individual devices, with their own individual processors, memory, buses, and failure modes. It is reasonable to view them as nodes with their own connections."

How do we account for skitter visits to router and host interfaces? Like many standard traceroute implementations, skitter sends three probe packets for each hop count. An IP address appears thus in a traceroute result if it appears in the replies to, at least, one of the three probes sent (but it may also appear two or three times). For each reply, we account one visit. If none of the three probes are returned, the hop is recorded as non-responding.

Even if an IP address is returned for a given hop count, it might not be valid. Due to the presence of poorly configured routers along traceroute paths, skitter occasionally records anomalies such as private IP addresses that are not globally routable. We account for invalid hops as if they were non-responding hops. The addresses that we consider as invalid are a subset of the special-use IPv4 addresses described in RFC 3330 [18]. Specifically, we eliminate visits to the private IP address blocks 10.0.0.0/8, 172.16.0.0/12, and 192.168.0.0/16. We also remove the loopback address block 127.0.0.0/8. In our data set, we find 4,435 different special addresses, more precisely 4,434 are private addresses and only one is a loopback address. Special addresses account for approximately 3% of the entire set of addresses seen in this trace. Though there were no visits in the data to the following address blocks, they too would be considered invalid: the "this network" block 0.0.0.0/8, the 6to4 relay any cast address block 192.88.99.0/24, the benchmark testing block 198.18.0.0/15, the multicast address block 224.0.0.0/4, and the reserved address block formerly known as the Class E addresses, 240.0.0.0/4, which includes the LAN broadcast address, 255.255.255.255.

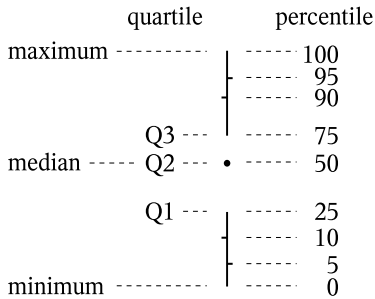

Figure 1: Quantiles key

## 3 Description of the Plots

In this report, we plot interface redundancy distributions. Since these distributions are generally skewed, quantile plots give us a better sense of the data than would plots of the mean and variance. There are several possible ways to calculate quantiles. We calculate them in the manner described by Jain [19, p. 194], which is: rounding to the nearest integer value to obtain the index of the element in question, and using the lower integer if the quantile falls exactly halfway between two integers.

Fig. 1 provides a key to reading the quantile plots found in subsequent sections of this report.

A dot marks the median (the $2^{nd}$ quartile, or $50^{th}$ percentile). The vertical line below the dot delineates the range from the minimum to the $1^{st}$ quartile, and leaves a space from the $1^{st}$ to the $2^{nd}$ quartile. The space above the dot runs from the $2^{nd}$ to the $3^{rd}$ quartile, and the line above that extends from the $3^{rd}$ quartile to the maximum. Small tick bars to either side of the lines mark some additional percentiles: bars to the left for the $10^{th}$ and $90^{th}$, and bars to the right for the $5^{th}$ and $95^{th}$.

In the case of highly skewed distributions, or distributions drawn from small amounts of data, the vertical lines or the spaces between them might not appear. For instance, if there are tick marks but no vertical line above the dot, this means that the $3^{rd}$ quartile is identical to the maximum value.

In the figures, each quantile plot sits directly above an accompanying bar chart that indicates the quantity of data upon which the quantiles were based. For each hop count, the bar chart displays the number of interfaces at that distance. For these bar charts, a log scale is used on the vertical axis. This allows us to identify quantiles that are based upon very few interfaces (fewer than twenty, for instance), and so for which the values risk being somewhat arbitrary.

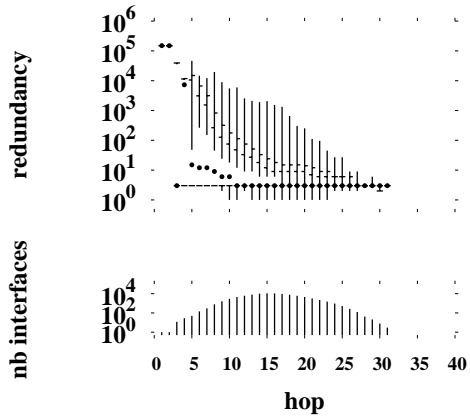

(a) arin

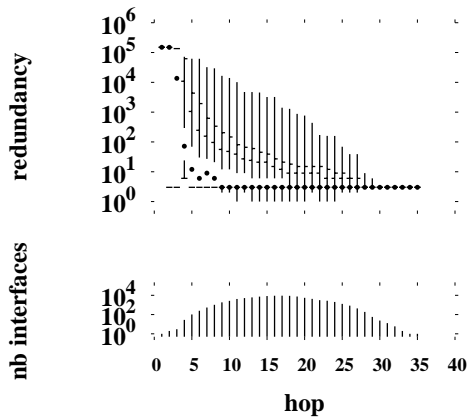

(b) champagne

Figure 2: Redundancy when probing with the pure forwards algorithm

## 4 Standard Traceroute

Our basis for comparison is the results from the standard forward tracing algorithm implemented in traceroute. All monitors operate from a set of common destinations, $D$. Each monitor probes forward starting from TTL=1 and increasing the TTL hop by hop towards each of the destinations in $D$ in turn. As it probes, a monitor $i$ updates the set, $S_i$, initially empty, of interfaces that it has visited.

Evaluating redundancy in the standard traceroute was already published in an authors' SIGMETRICS 2005 paper [10]. For comparison reasons in the next sections of this report, we summarize in this section our redundancy evaluation of standard traceroute.

Fig. 2 shows redundancy distributions for two skitter monitors: `arin` and `champagne`. The results presented in Fig. 2 are representative of all the skitter monitors. Interested readers might find plots for the 22 other skitter monitors in our technical report [20].

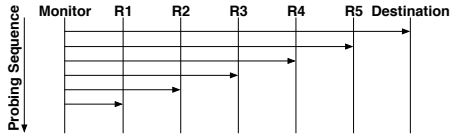

Figure 3: Pure backwards algorithm

Looking first at the histograms for interface counts (the lower half of each plot), we find data consistent with distributions typically seen in such cases. If we were to look at a plot on a linear scale (not shown here), we would see that these distributions display the familiar bell-shaped curve typical of internet interface distance distributions [21]. If we concentrate on `champagne`, we see that it discovers 92,354 unique and valid IP addresses. The interface distances are distributed with a mean at 17 hops corresponding to a peak of 9,135 interfaces that are visited at that distance.

The quantile plots show the nature of the redundancy problem. Looking at how the redundancy varies by distance, we see that the problem is worse the closer one is to the monitor. This is what we expect given the tree-like structure of routing from a monitor, but here we see how serious the phenomenon is from a quantitative standpoint. For the first two hops from each monitor, the median redundancy is 150,000. A look at the histograms shows that there are very few interfaces at these distances. Just one interface for `arin`, and two ($2^{nd}$ hop) or three ($3^{rd}$ hop) for `champagne`. These close to the monitor interfaces are only visited three times, as represented by the presence of the $5^{th}$ and $10^{th}$ percentile marks (since there are only two data points, the lower values point is represented by the entire lower quarter of values on the plot).

Beyond three hops, the median redundancy drops rapidly. By the eleventh hop, in both cases, the median is below ten. However, the distributions remain highly skewed. Even fifteen hops out, some interfaces experience a redundancy on the order of several hundred visits. With small variations, these patterns are repeated for each of the monitors.

## 5 Backward Tracing

As seen and discussed in Sec. 4, the most worrisome feature of redundancy in a standard measurement system is the exceptionally high number of visits to the median interfaces close in to the monitor. Also of concern is the heavy tail of the distribution at more distant hop counts, with a certain number of interfaces consistently receiving a high number of visits. Our approach here is to tackle the first problem head-on, and then to see if the second problem remains.

The large number of visits to nodes close in to a monitor is easily explained by the tree-like or conal structure of the graph generated by traceroutes from a single monitor, as described by Broido and claffy [17]. There are typically only a few interfaces close to a monitor, and these interfaces must therefore be visited by a large portion of the traceroutes. The solution to this problem is simple, at least in principle: these close in interfaces must be skipped most of the time.

Traceroute works forward from source to destination. Its first set of probes goes just one hop, its second set goes two, and so forth. It would seem that the best way to reduce intra-monitor redundancy is to start further out and probes backward, i.e., decreasing TTLs. Govindan and Tangmunarunkit [11] do just this in the Mercator system. Using a probing strategy based upon IP address prefixes, Mercator conducts a check before probing the path to a new address that has a prefix $P$. If paths to an address in $P$ already exist in its database, Mercator starts probing at the highest hop count for a responding router seen on those paths. No results have been published on the performance of this heuristic, though it seems to us an entirely reasonable approach in light of our data.

The Mercator heuristic requires that a guess be made about the relevant prefix length for an address. That guess is based upon the class that the address would have had before the advent of classless inter-domain routing (CIDR) [22]. In this technical report, we have tested a number of simple heuristics that do not require us to hazard such a guess.

Our algorithms work backwards. As illustrated in Fig. 3, a monitor sends its first probes to the destination, its second to one hop short of the destination, and so forth. Now arises the question of when to stop backward probing. Based on the tree-like structure of routes emanating from a single point (i.e., the traceroute source), we choose a stopping rule based on the set of interfaces previously encountered. A monitor will stop backward probing when an already visited interface is encountered. The only redundancy such a strategy should produce would be on interfaces that are branching points in paths between a monitor and its destinations. A backward probing scheme uses the set, $S_i$, of interfaces that a monitor $i$ has visited. In early probing, $S_i$ will have few elements, and so paths should be traced from the destination almost all the way back to the monitor. Later probing should terminate further and further out, as more and more interfaces are added to $S_i$.

There are practical problems with a strategy of back-wards probing. They arise because of inherent flaws with methods for establishing the number of hops from a monitor to a destination. These methods rely upon the sending of a ping packet (or a *scout* packet, following Moors' terminology [12]), and the examination of the time to live (TTL) value in the IP packet that the destination returns. Various heuristics have been described, by Templeton and Levitt [23], Jin et al. [24], Moors [12] and Beverly [25], to guess the original TTL (typically one of a few standard values) based upon the observed value, and thus to guess the hop count from destination to monitor. While these heuristics have been shown to work well, the most serious problem is that they cannot work when the destination does not reply, as is often the case (40.3% in our data). In such a case, a system that takes a backwards probing approach will ideally start from the most distant interface that responds with an ICMP "TTL expired" packet when discarding a hop-limited probe. In practice, this might take some search to locate, adding redundancy.

Furthermore, as established by Paxson [26] based upon data from 1995, and confirmed with data from 2002 by Amini et al. [27], a considerable number of paths in the Internet are asymmetric: most recently almost 70%. This is a less serious problem, however, as the differences in routing often do not translate into considerable differences in hop count. Paxson's work indicated that differences in one or two hops were typical. For the purposes of our simulations we assume that, if a destination does reply to a ping, the system thereby learns the correct number of hops on the forward path.

## 5.1 Pure Backwards

We simulate an algorithm for backwards probing in which the most distant responding interface is assumed to be known a priori. Called *pure backward probing*, this algorithm is unrealistic because of its assumption. However, its performance sets a benchmark. Against that algorithm, we later compare algorithms that use only information that is actually available to a monitor.

Fig. 4 shows redundancy for monitors running the pure backwards algorithm[1]. We notice a significant drop in comparison to the redundancy in straightforward tracerouting shown in Fig. 2. The median drops for the close interfaces, and the distribution tails are significantly shortened overall. However, Figs. 4(a) and 4(b) show still high redundancy for interfaces located one hop from the traceroute source.

We hypothesize that these remanent high redundancies close to the monitors are caused by the existence of firewalls or gateways that either do not permit probes to pass through them, or do not permit replies to return.

---

[1]Plots for others monitors can be found in an appendix at the end of this report.

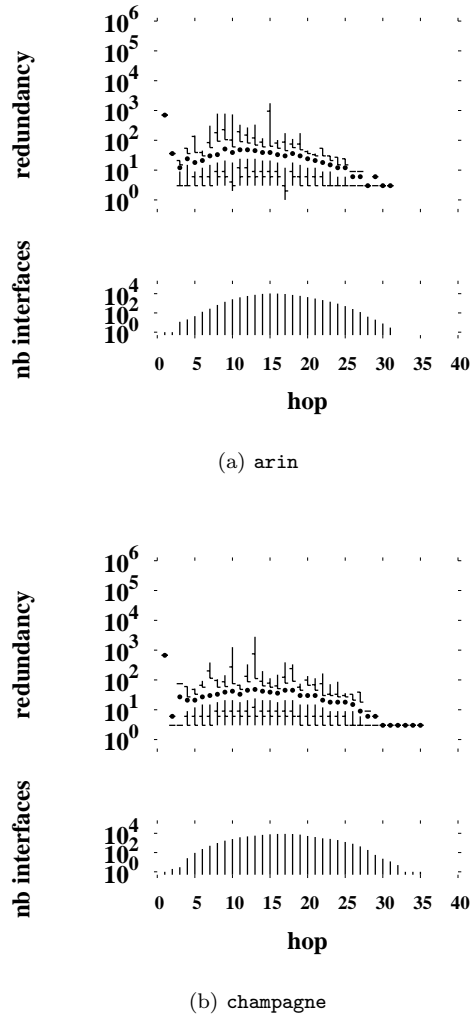

(a) `arin`

(b) `champagne`

Figure 4: Redundancy when probing with the pure backwards algorithm

If the destination addresses are invalid, these interfaces could also be default free routers. Under pure backwards probing, a node situated immediately in front of such a device, whatever it might be, will be visited again and again, for each destination that lies beyond, thus resulting in a high visit count for one of its interfaces. Without any further knowledge, the actual cause of such high redundancy under backwards probing remains for us an open question.

However, Figs. 4(a) and 4(b) show that maximum redundancies are in the thousands, rather that the hundreds of thousands as before. Furthermore, median values are a little higher than with standard traceroute. The strong drop in redundancy close to the monitor thus comes at the expense of some increased redundancy further out. The overall effect is one of smoothing the load.

There are costs associated with this drop in redundancy. We measure these in terms of the number of interfaces missed, using the set $S_i$ of interfaces visited by the standard algorithm as the reference. For the pure backwards algorithm, the numbers missed are relatively small, as shown in Table 7. Table 7 gives also the cost in term of links missed by the pure backwards algorithm.

Interfaces will necessarily be missed in backwards probing when a hard and fast rule is applied that requires probing to stop once an already visited interface is encountered. Any routing change that might have taken place between the monitor and that interface will go unnoticed. A routing system that adopts a backwards probing algorithm should also adopt a strategy for periodically reprobing certain paths, so as not to miss such changes entirely. So long as the portion of interfaces missed is small, we believe that the development of such a reprobing strategy can be left to future work.

## 5.2 Ordinary Backwards

The *ordinary backwards* algorithm works in much the same way as perfect backwards, but it is a more realistic algorithm. Just as with pure backwards, when a destination responds, the monitor starts probing backwards from the destination until an already visited interface is met. However, when a destination doesn't reply, the monitor, since it cannot know a priori the most distant responding interface along the path, gives up probing for this particular destination altogether. This is the first of two building blocks that will be used by the algorithm presented in Sec. 5.4, and is not intended to be used in isolation.

Approximately 40% the traceroutes in our data set terminate in a non-responding destination. What does this mean in terms of interfaces that are missed? Table 2 shows the costs of not probing these paths combined with the early stopping that is in any case associated with backwards probing. What is remarkable to note is that, compared to perfect backwards probing, ordinary backwards probing only misses an additional 16% of interfaces.

Fig. 5 shows trends very similar to those observed with perfect backwards probing, but some high values are no longer present.

## 5.3 Searching

If we are to use ordinary backwards probing as one element of a larger probing strategy, we need a second element to handle destinations that do not respond. Since the last responding interface on a path to such a destination cannot be known a priori, the monitor must search for it. The search cost is what will make the difference with respect to the pure backwards algorithm.

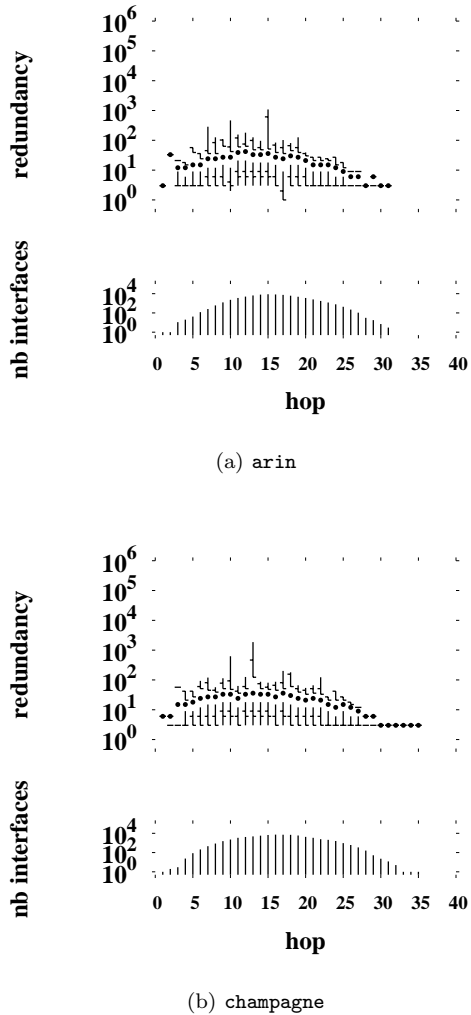

(a) `arin`

(b) `champagne`

Figure 5: Redundancy when probing with the ordinary backwards algorithm

Our algorithm, labeled *searching*, now sends its initial probe with a TTL value $h$. If it receives a response, it continues to probe forwards, to TTLs $h + 1$, $h + 2$, and so forth. When the farthest responding interface is found, probing resumes from TTL $h - 1$, and probes backwards, to TTLs $h - 2$, $h - 3$, and so back. If, at any point, a monitor $i$ visits an interface that is in its set $S_i$ of interfaces already viewed, probing for that destination stops. The working of the searching algorithm is illustrated in Fig. 6, where $h = 3$ and $R_5$ being the last responding interface.

If the algorithm is supposed to start probing from a midpoint $h$ in the network, we have to decide which value give to $h$. Doubletree [10], proposed by the au-thors of this report, is a cooperative and efficient algorithm for large-scale topology discovery. Each Doubletree monitor starts probing at some hop $h$ from itself, performing forwards probing from $h$ and backwards probing from $h - 1$. The value $h$ is fixed by each monitor according to its probability $p$ of hitting a destination with the very first probe sent. This choice is driven by the risk of probing looking like a distributed denial-of-service (DDoS) attack. Indeed, when probes sent by multiples monitors converge towards a given destination, the probe traffic might appear, for an end-host, as a DDoS attack. Doubletree aims to minimize this risk and, therefore, each monitor chooses an appropriate $h$ value.

| Monitor | Interfaces | | | Links | | |
|---|---|---|---|---|---|---|
| | total | discovered | % missed | total | discovered | % missed |
| arin | 92,381 | 88,204 | 4.52% | 101,850 | 92,602 | 9.09% |
| champagne | 92,354 | 88,012 | 4.70% | 101,652 | 92,331 | 9.17% |

Table 1: Interfaces missed by the pure backwards algorithm

| Monitor | Interfaces | | | Links | | |
|---|---|---|---|---|---|---|
| | total | discovered | % missed | total | discovered | % missed |
| arin | 92,381 | 73,529 | 20.40% | 101,850 | 75,163 | 27.21% |
| champagne | 92,354 | 73,410 | 20.51% | 101,652 | 74,987 | 27.24% |

Table 2: Interfaces missed by the ordinary backwards algorithm

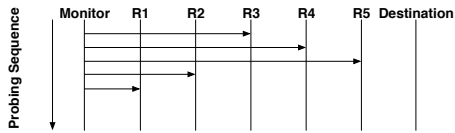

Figure 6: Searching algorithm with $h = 3$

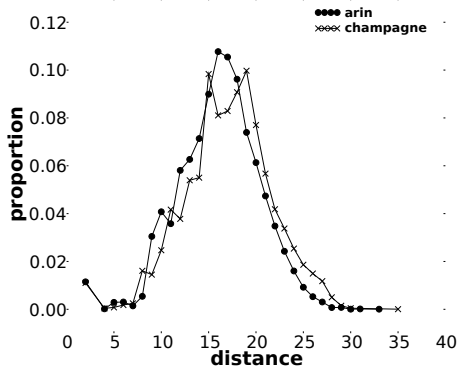

Figure 7: Incomplete paths distribution

In this report, we are not interested in large-scale distributed probing, i.e., from a large set of monitors that cooperate when probing towards a large set of destinations. We consider that each monitor works in isolation of others. It does not make sense to choose the $h$ value like Doubletree does.

Fig. 7 shows the incomplete paths distribution, i.e., the distance distribution of the last responding hop when a traceroute does not terminate by hitting the destination. Such a case occurs in approximately 40% of the traceroutes in our dataset. We propose that each monitor tunes its $h$ value with the mean hop count for its incomplete traceroutes. A monitor can easily evaluates its own $h$ value by performing a small number of standard traceroutes.

In the special case where there is no response at distance $h$, the distance is halved, and halved again until there is a reply, and probing continues forwards and backwards from that point.

Our results in Fig. 8 indicate that low redundancy can be achieved. We tested the heuristic algorithm using only those traceroutes for which the destination does not respond.

However, we notice that close to the monitor, in the fashion of the pure backwards algorithm (see Fig. 4), the redundancy is still high. We believe that this is caused by very short paths for which the last responding interface is close to the monitor. For those paths, there is a high probability that sending the first probe at $h$ hops to the monitor will corresponds a non-response. The $h$ value will be divided by two, again and again, until reaching a responding interface that will be located close to the monitor, increasing therefore the redundancy of such interfaces.

## 5.4 Searching Ordinary Backwards

In this section, we study a comprehensive strategy for reducing probing redundancy. We employ ordinary backwards probing, along with the heuristic algorithm for cases in which the destination does not respond. This algorithm is called *searching ordinary backwards*.

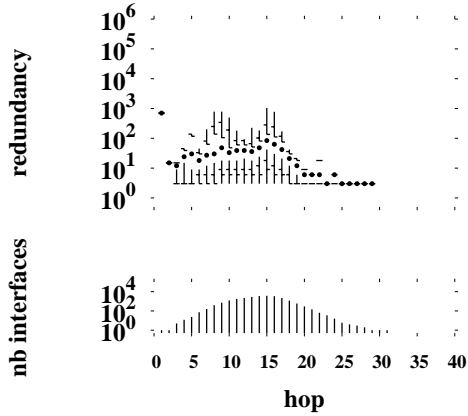

(a) `arin`

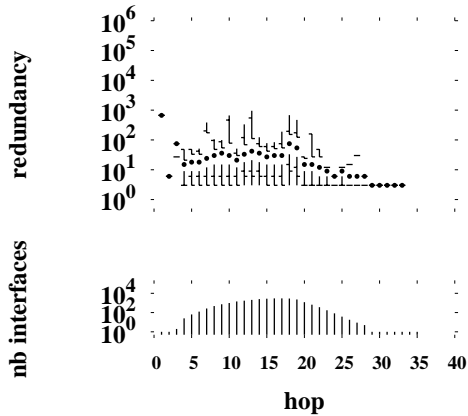

(b) `champagne`

Figure 8: Redundancy when probing with the searching algorithm

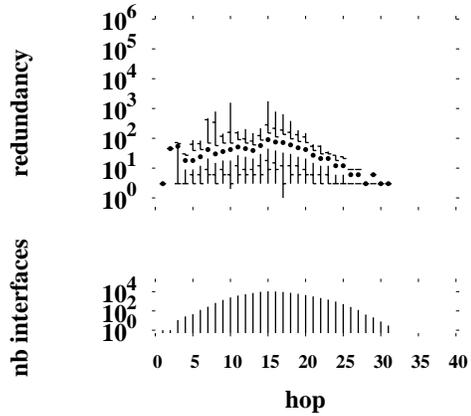

(a) `arin`

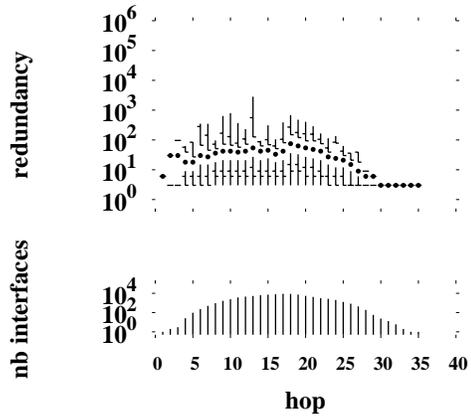

(b) `champagne`

Figure 9: Redundancy when probing with the searching ordinary backwards algorithm

Fig. 9 shows redundancy reduction similar to that obtained with the other algorithms examined so far.

Table 3 shows the interfaces and links missed when probing with the searching ordinary backwards algorithm. Table 3 indicates that the numbers of missed interfaces are a bit smaller (this is specially true for `arin`) than with the supposedly pure backwards algorithm, a surprising fact which will be explained in the next section.

# 6 Algorithms Comparison

Fig. 10 shows the trade-offs between redundancy and missed address interfaces. Redundancy is here represented by the mean number of visits per valid discovered interface, and the missed addresses are expressed as a proportion, using the standard traceroute as the reference. Results shown in Fig. 10 are representative for all monitors, as demonstrated by Table 4 giving the mean over the 24 monitors.

Our goal is to avoid both redundancy and missed interfaces as much as possible, however there is a tradeoff between the two. Extremes are represented by the standard traceroute, which by definition misses no interfaces, but has high average redundancy, and by the heuristic algorithm which, having been applied to just those traceroutes for which the destination did not respond, necessarily misses a large number of interfaces.

We see that the ordinary backwards provides the lowest redundancy but, as it is applied on only respond-ing traceroutes, it misses a lot of interfaces. The most interesting comparison is between the pure backwards algorithm and searching ordinary backwards. Both are based upon the full set of traceroutes, and so are strictly comparable. The searching ordinary backwards algorithm manages to outperform the pure backwards algorithm by paying a slight price in terms of increased redundancy (in the case of arin).

| Monitor | Interfaces | | | Links | | |
|---|---|---|---|---|---|---|
| | total | discovered | % missed | total | discovered | % missed |
| arin | 92,381 | 90,156 | 2.40% | 101,850 | 94,149 | 7.57% |
| champagne | 92,354 | 87,946 | 4.77% | 101,652 | 91,123 | 10.36% |

Table 3: Interfaces missed by the searching ordinary backwards algorithm

| Algorithms | Mean visit | Prop. missed |
|---|---|---|
| standard | 25.08 | 0 |
| heuristic | 9.16 | 0.74 |
| search. ord. bwd | 6.21 | 0.03 |
| pure bwd | 5.58 | 0.04 |
| ordinary bwd | 4 | 0.2 |

Table 4: Algorithms comparison - mean

# 7 Conclusion

This technical report addresses the area of efficient probing in the context of traceroute monitors working in isolation from each others. Prior work stated that standard traceroutes are particularly inefficient by repeatedly reprobing the same interfaces close to the monitor. The solution to this redundancy problem is, at least in principle, simple: those interfaces close to the monitor must be skipped most of the time. It seems that the best way to achieve this solution is to probe backwards from the destinations and stop when encountering a previously seen interface.

In this report, we perform the first careful study of the efficiency of backwards probing, by evaluating it in terms of redundancy reduction and information lost.

Nevertheless, we state that a pure backwards probing algorithm is unrealistic as it is based on the assumption that destinations reply to probes. We therefore propose an algorithm that searches for the last responding interface. The key idea of this algorithm is to start probing at some hop $h$ from the monitor, probe forwards from $h$ until the last responding interface and, then, probe backwards from $h - 1$ until reaching an already discovered interface.

We finally propose a realistic algorithm that can handle both cases, i.e., responding and non-responding des-tinations. We evaluate this algorithm and state that it is capable of reducing probe traffic by a factor of 10, while only missing 4% of the interfaces discovered by a standard traceroute.

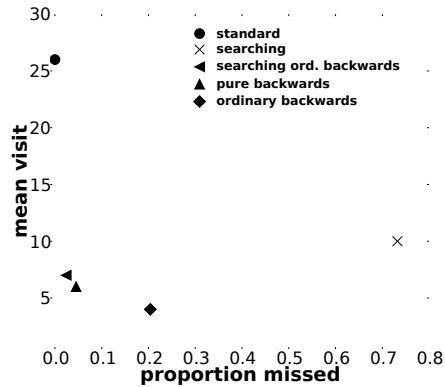

(a) arin

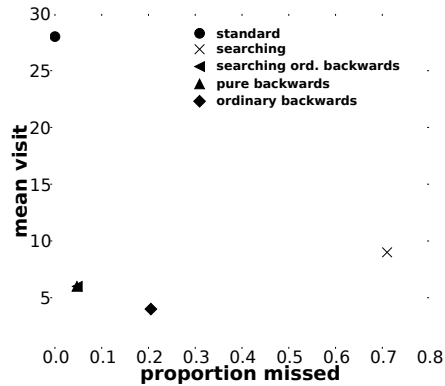

(b) champagne

Figure 10: Algorithms comparison

As a future work, we aim to provide to the research community an implementation of the algorithms discussed in this report.

# References

[1] V. Jacobsen et al., "traceroute," UNIX, man page, 1989, see source code: ftp://ftp.ee.lbl.gov/traceroute.tar.gz.

[2] B. Huffaker, D. Plummer, D. Moore, and k. claffy, "Topology discovery by active probing," in *Proc. Symposium on Applications and the Internet*, Jan. 2002.

[3] "IPv6 scamper," WAND Network Research Group. Web site: http://www.wand.net.nz/ mjl12/ipv6-scamper/.

[4] Y. Shavitt and E. Shir, "DIMES: Let the internet measure itself," *ACM SIGCOMM Computer Communication Review*, vol. 35, no. 5, 2005, http://www.netdimes.org.

[5] D. P. Anderson, J. Cobb, E. Korpela, M. Lebofsky, and D. Werthimer, "SETI@home: An experiment in public-resource computing," *Communications of the ACM*, vol. 45, no. 11, 2002, see http://setiathome.berkeley.edu/.

[6] F. Georgatos, F. Gruber, D. Karrenberg, M. Santcroos, A. Susanj, H. Uijterwaal, and R. Wilhelm, "Providing active measurements as a regular service for ISPs," in *Proc. Passive and Active Measurement (PAM) Workshop*, Apr. 2001.

[7] A. McGregor, H.-W. Braun, and J. Brown, "The NLANR network analysis infrastructure," *IEEE Communications Magazine*, vol. 38, no. 5, 2000.

[8] N. Spring, D. Wetherall, and T. Anderson, "Scriptroute: A public internet measurement facility," in *Proc. 4th USENIX Symposium on Internet Technologies and Systems*, Mar. 2002, see also http://www.cs.washington.edu/research/networking/scriptroute/.

[9] PlanetLab Consortium, "PlanetLab project," 2002, see http://www.planet-lab.org.

[10] B. Donnet, P. Raoult, T. Friedman, and M. Crovella, "Efficient algorithms for large-scale topology discovery," in *Proc. ACM SIGMETRICS*, 2005, see http://trhome.sourceforge.net.

[11] R. Govindan and H. Tangmunarunkit, "Heuristics for internet map discovery," in *Proc. IEEE INFOCOM*, Mar. 2000.

[12] T. Moors, "Streamlining traceroute by estimating path lengths," in *Proc. IEEE International Workshop on IP Operations and Management (IPOM)*, Oct. 2004.

[13] J. J. Pansiot and D. Grad, "On routes and multicast trees in the internet," *ACM SIGCOMM Computer Communication Review*, vol. 28, no. 1, pp. 41–50, Jan. 1998.

[14] K. Keys, "iffinder," a tool for mapping interfaces to routers. See http://www.caida.org/tools/measurement/iffinder/.

[15] N. Spring, R. Mahajan, and D. Wetherall, "Measuring ISP topologies with Rocketfuel," in *Proc. ACM SIGCOMM*, Aug. 2002.

[16] R. Teixeira, K. Marzullo, S. Savage, and G. Voelker, "In search of path diversity in ISP networks," in *Proc. ACM SIGCOMM Internet Measurement Conference (IMC)*, Oct. 2003.

[17] A. Broido and k. claffy, "Internet topology: Connectivity of IP graphs," in *Proc. SPIE International Symposium on Convergence of IT and Communication*, Aug. 2001.

[18] IANA, "Special-use IPv4 addresses," Internet Engineering Task Force, RFC 3330, 2002.

[19] R. K. Jain, *The Art of Computer Systems Performance Analysis*. John Wiley, 1991.

[20] B. Donnet, P. Raoult, T. Friedman, and M. Crovella, "Efficient algorithms for large-scale topology discovery," arXiv, cs.NI 0411013 v1, Nov. 2004.

[21] F. Begtasevic and P. Van Mieghem, "Measurement of the hopcount in the internet," in *Proc. Passive and Active Measurement Workshop (PAM)*, Apr. 2001.

[22] V. Fuller, T. Li, J. Yu, and K. Varadhan, "Classless inter-domain routing (CIDR): an address assignment and aggregation strategy," Internet Engineering Task Force, RFC 1519, Sep. 1993.

[23] S. J. Templeton and K. E. Levitt, "Detecting spoofed packets," in *Proc. Third DARPA Information Survivability Conference and Exposition (DISCEX III)*, Apr. 2003.

[24] C. Jin, H. Wang, and K. G. Shin, "Hop-count filtering: An effective defense against spoofed DDoS traffic," in *Proc. ACM Conference on Computer and Communication Security (CCS)*, Oct. 2003.

[25] R. Beverly, "A robust classifier for passive TCP/IP fingerprinting," in *Proc. Passive and Active Measurement Workshop (PAM)*, Apr. 2004.

[26] V. Paxson, "End-to-end routing behavior in the internet," in *Proc. ACM SIGCOMM*, Aug. 1996.
10

# A Pure Backwards

## A.1 Redundancy Reduction

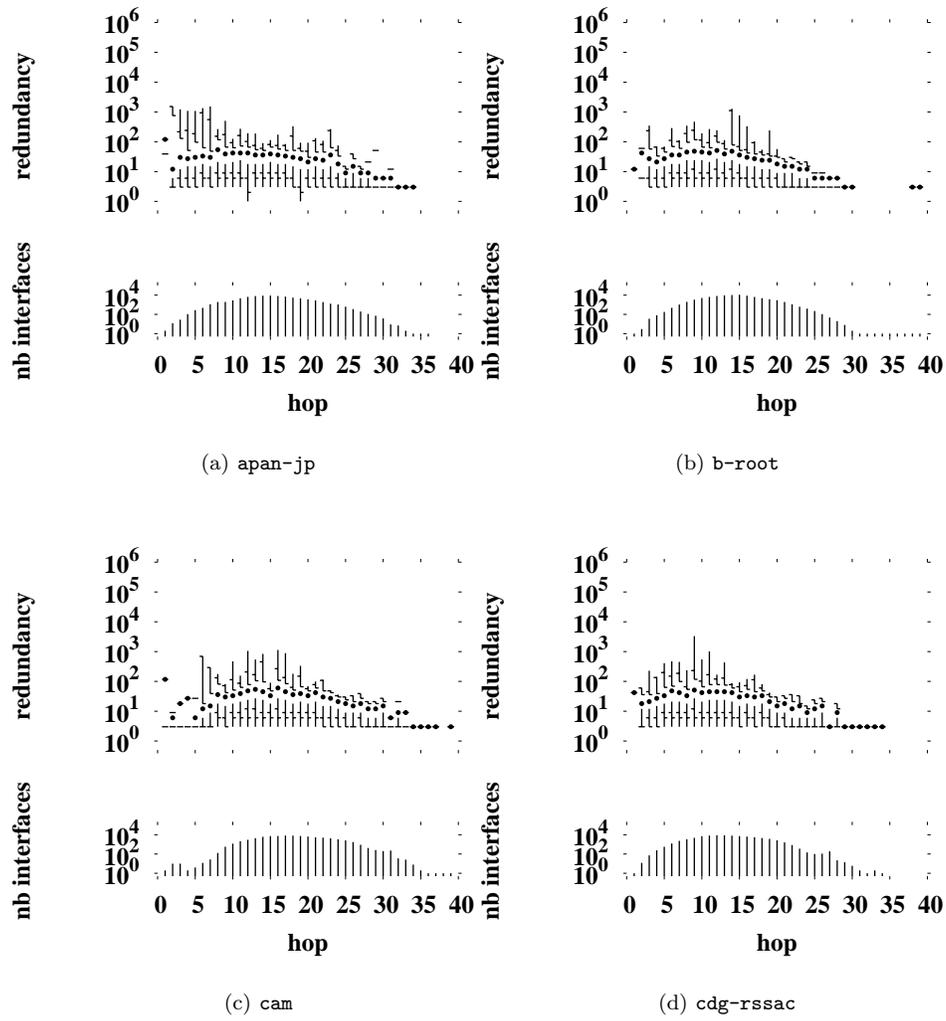

Figure 11: Redundancy when probing with the pure backwards algorithm - 1

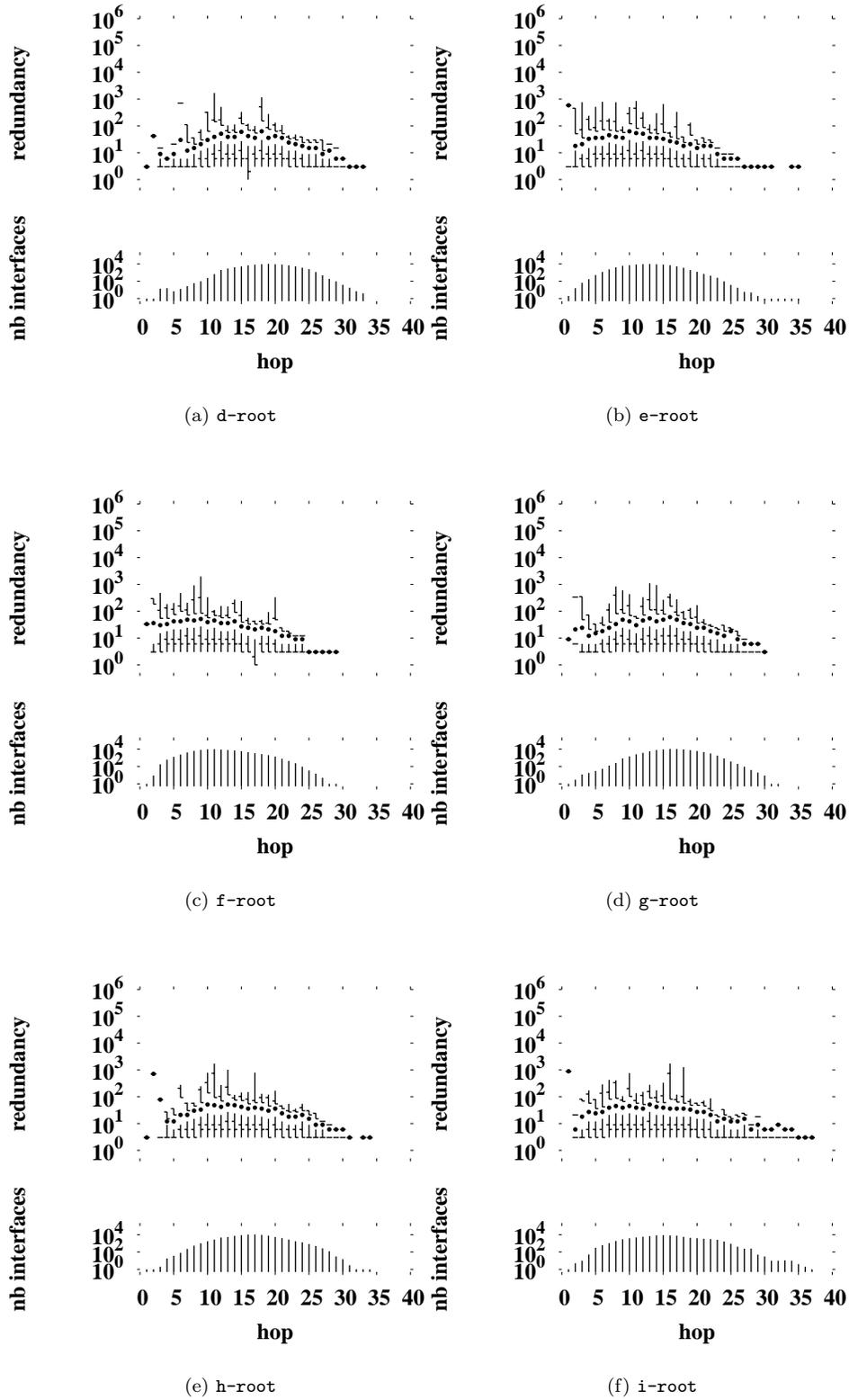

Figure 12: Redundancy when probing with the pure backwards algorithm - 2

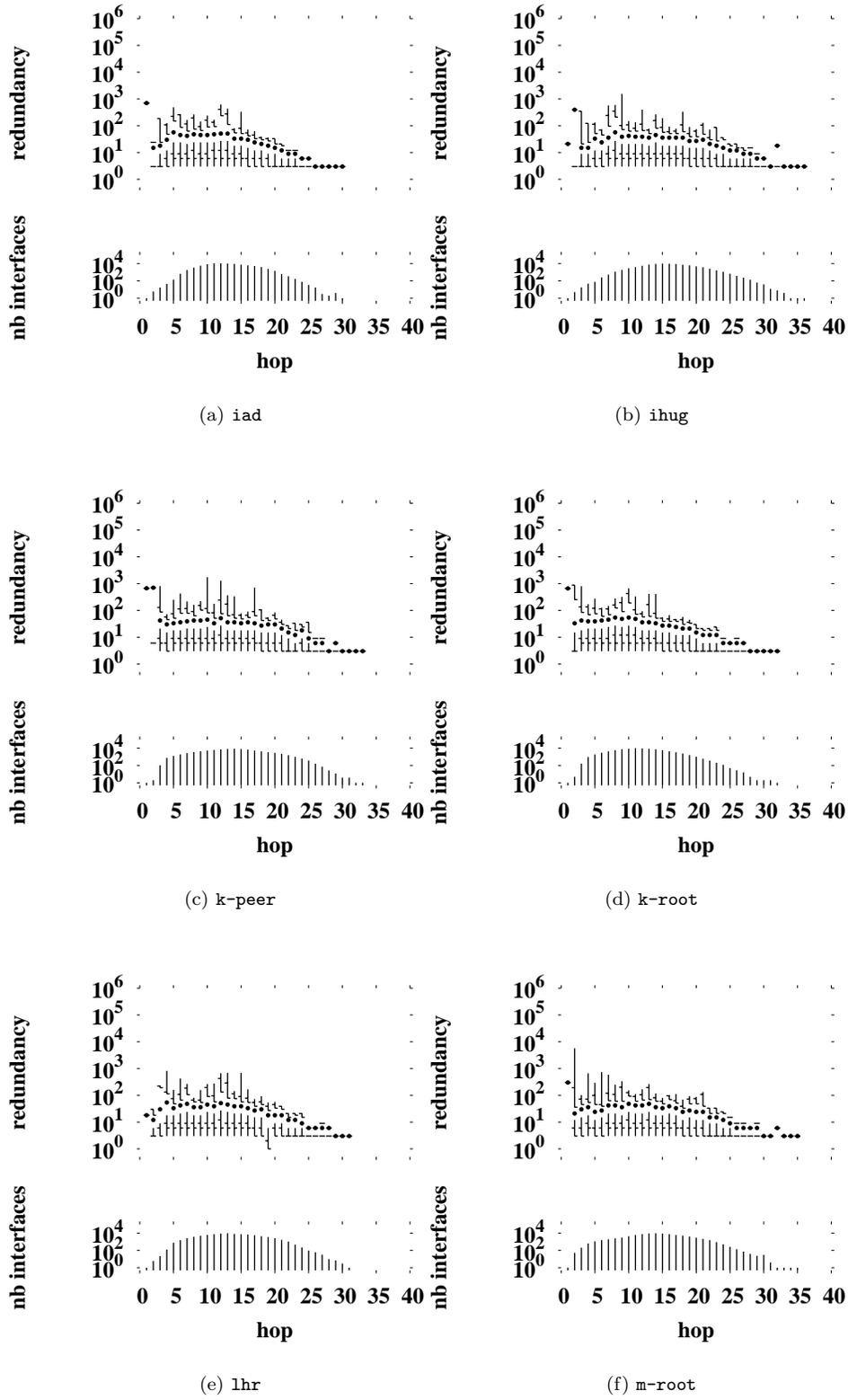

Figure 13: Redundancy when probing with the pure backwards algorithm - 3

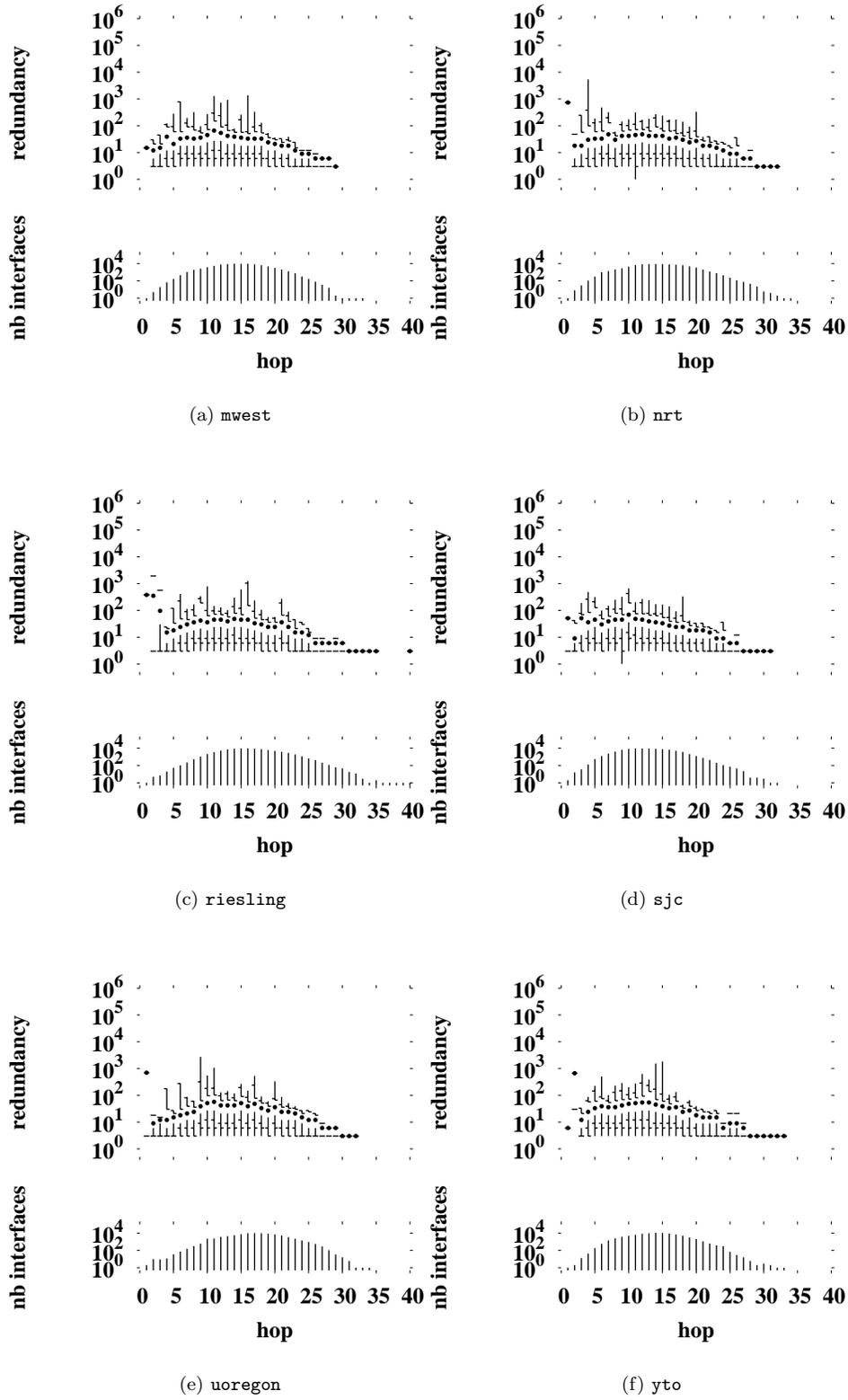

Figure 14: Redundancy when probing with the pure backwards algorithm - 4

## A.2 Losses

| Monitor | Interfaces | | | Links | | |
|---|---|---|---|---|---|---|
| | total | discovered | % missed | total | discovered | % missed |
| `apan-jp` | 86,763 | 82,659 | 0.04730% | 96,908 | 88,562 | 0.08612% |
| `b-root` | 92,754 | 88,565 | 0.04516% | 103,595 | 94,379 | 0.08896% |
| `cam` | 90,796 | 86,799 | 0.04402% | 101,068 | 92,546 | 0.08431% |
| `cdg-rssac` | 90,962 | 87,015 | 0.04339% | 100,258 | 92,114 | 0.08123% |
| `d-root` | 91,136 | 87,228 | 0.04288% | 100,821 | 92,757 | 0.07998% |
| `e-root` | 90,952 | 86,592 | 0.04793% | 102,749 | 93,093 | 0.09397% |
| `f-root` | 92,123 | 88,136 | 0.04327% | 101,956 | 93,058 | 0.08727% |
| `g-root` | 91,547 | 87,108 | 0.04848% | 103,872 | 93,742 | 0.09752% |
| `h-root` | 91,825 | 87,725 | 0.04465% | 102,948 | 94,106 | 0.08588% |
| `i-root` | 91,942 | 87,827 | 0.04475% | 104,017 | 94,657 | 0.08998% |
| `iad` | 92,175 | 88,092 | 0.04429% | 102,324 | 93,046 | 0.09067% |
| `ihug` | 94,719 | 89,715 | 0.05282% | 107,979 | 96,292 | 0.10823% |
| `k-peer` | 91,851 | 87,730 | 0.04486% | 103,672 | 94,353 | 0.08988% |
| `k-root` | 91,726 | 87,806 | 0.04273% | 101,974 | 93,858 | 0.07958% |
| `lhr` | 92,079 | 88,215 | 0.04196% | 101,188 | 92,837 | 0.08252% |
| `m-root` | 92,347 | 88,078 | 0.04622% | 101,321 | 92,247 | 0.08955% |
| `mwest` | 91,525 | 87,388 | 0.04520% | 103,074 | 93,881 | 0.08918% |
| `nrt` | 92,021 | 87,897 | 0.04481% | 101,286 | 92,276 | 0.08895% |
| `riesling` | 90,913 | 86,766 | 0.04561% | 100,426 | 91,334 | 0.09053% |
| `sjc` | 91,459 | 87,433 | 0.04401% | 101,665 | 92,695 | 0.08823% |
| `uoregon` | 90,585 | 86,624 | 0.04372% | 100,851 | 92,360 | 0.08419% |
| `yto` | 91,200 | 87,199 | 0.04387% | 102,625 | 93,766 | 0.08632% |

Table 5: Interfaces missed by the pure backwards algorithm

# B Ordinary Backwards

## B.1 Redundancy Reduction

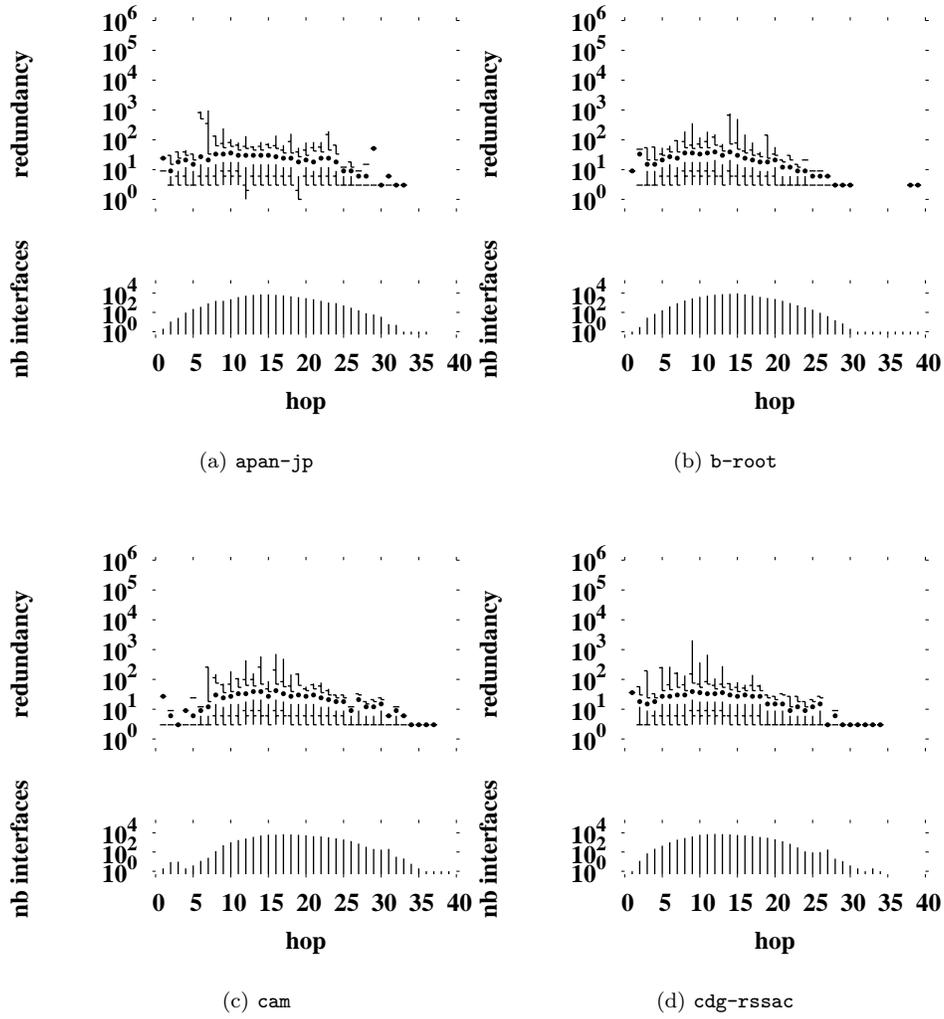

Figure 15: Redundancy when probing with the ordinary backwards algorithm - 1

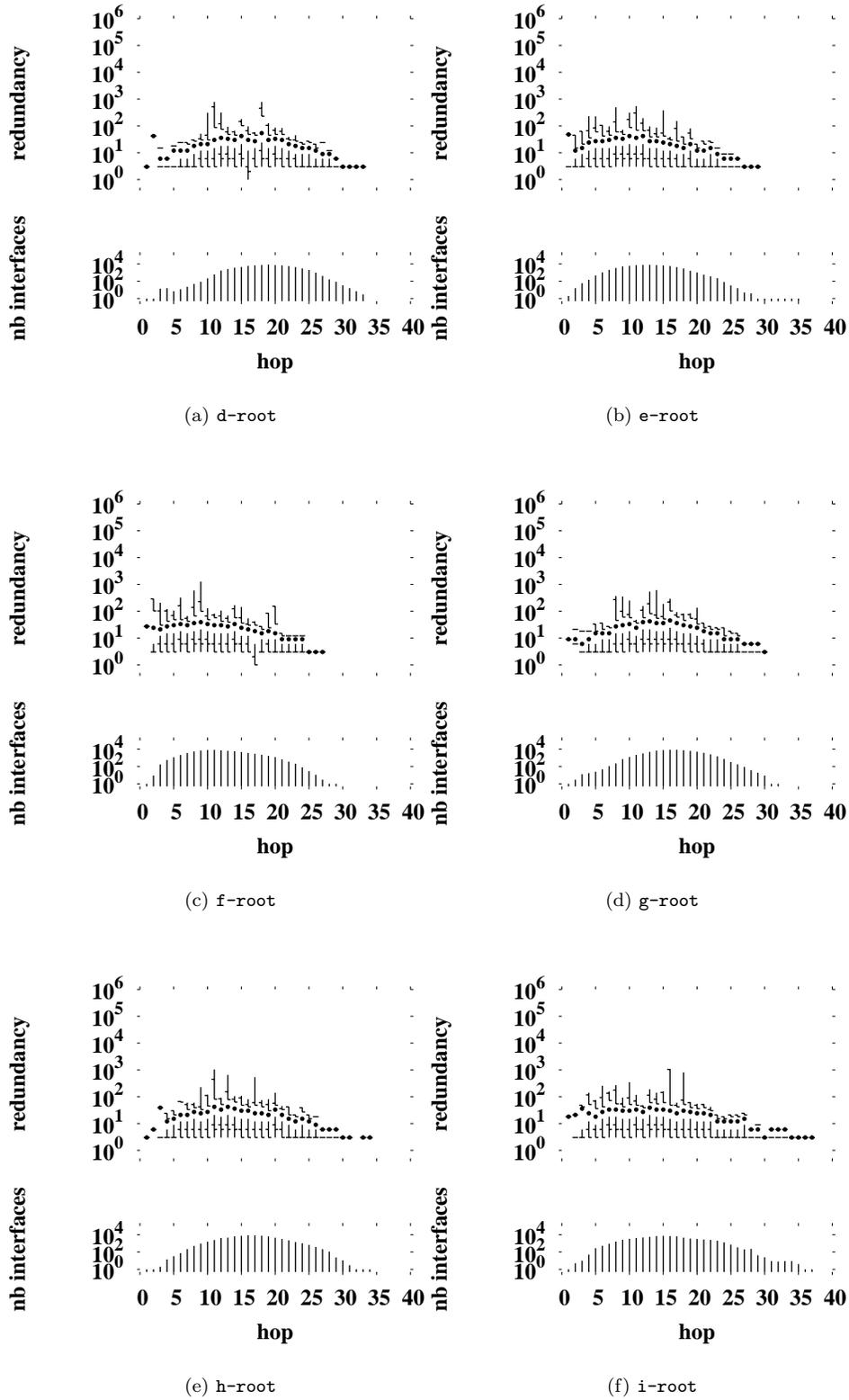

Figure 16: Redundancy when probing with the ordinary backwards algorithm - 2

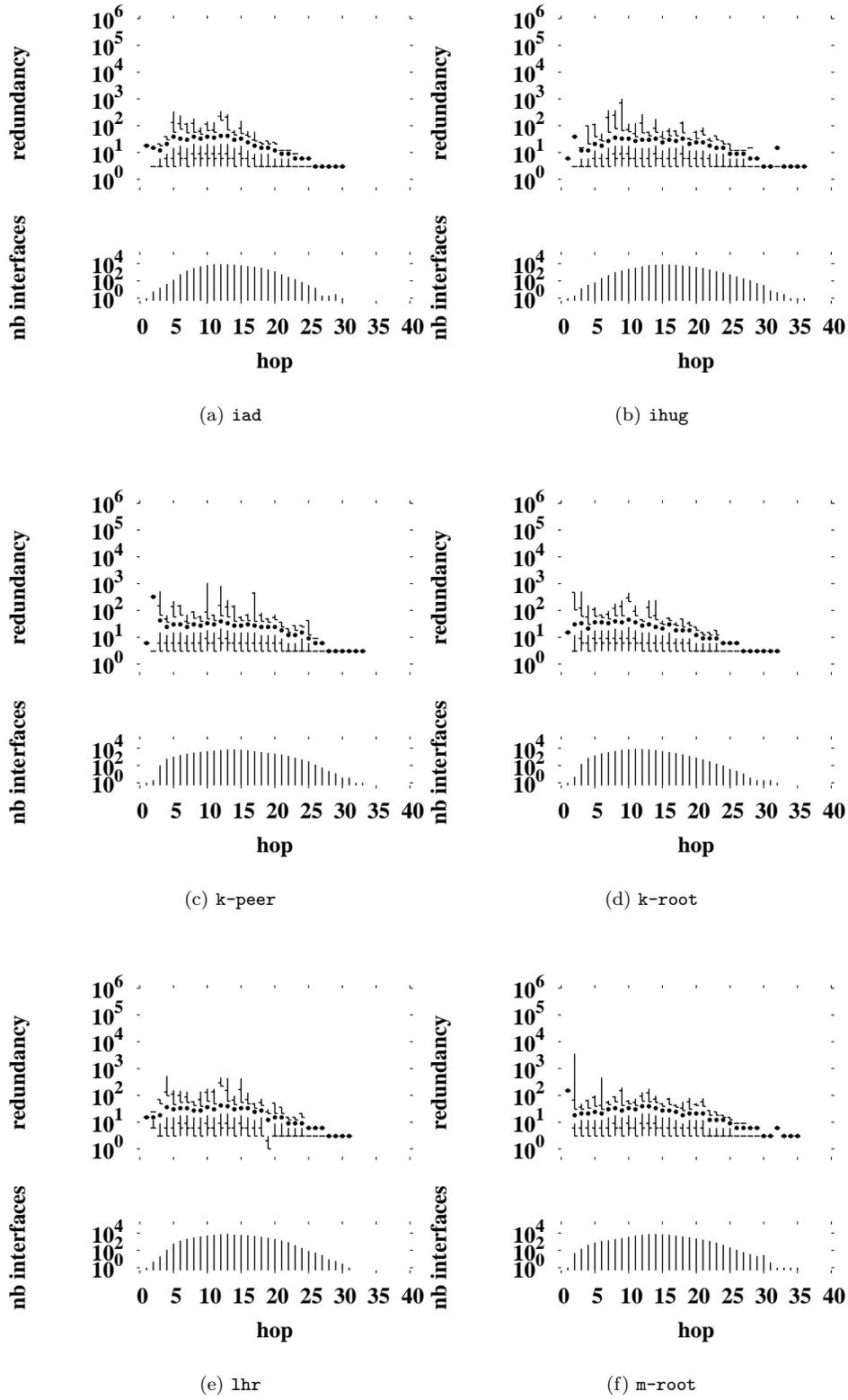

Figure 17: Redundancy when probing with the ordinary backwards algorithm - 3

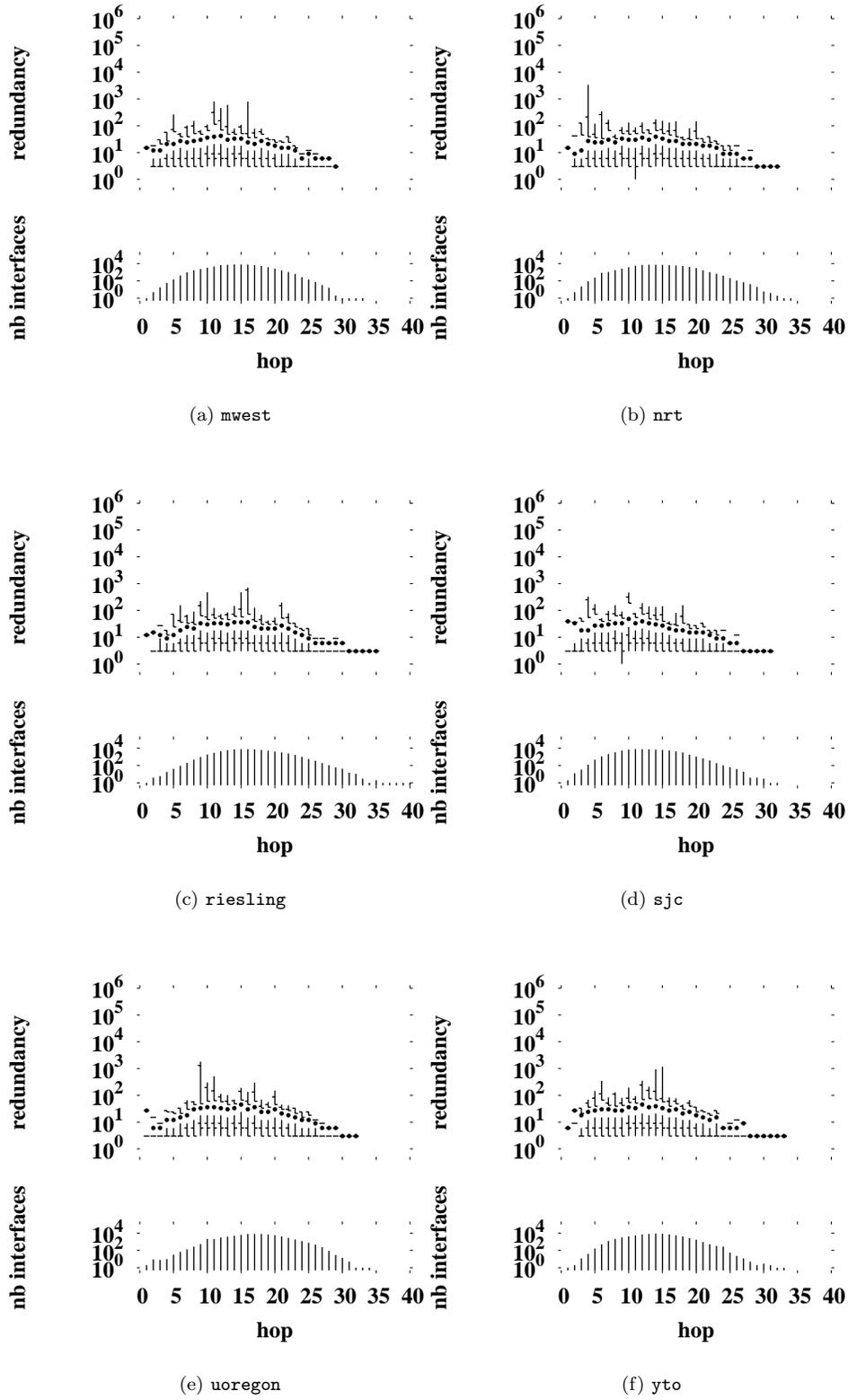

Figure 18: Redundancy when probing with the ordinary backwards algorithm - 4

## B.2 Losses

| Monitor | Interfaces | | | Links | | |
|---|---|---|---|---|---|---|
| | total | discovered | % missed | total | discovered | % missed |
| apan-jp | 86,763 | 68,105 | 0.21504% | 96,908 | 70796 | 0,26945% |
| b-root | 92,754 | 73,673 | 0.20571% | 103,595 | 76701 | 0,25960% |
| cam | 90,796 | 72,239 | 0.20438% | 101,068 | 75367 | 0,25429% |
| cdg-rssac | 90,962 | 72,345 | 0.20466% | 100,258 | 74897 | 0,25295% |
| d-root | 91,136 | 72,708 | 0.20220% | 100,821 | 75669 | 0,24947% |
| e-root | 90,952 | 71,967 | 0.20873% | 102,749 | 75477 | 0,26542% |
| f-root | 92,123 | 73,389 | 0.20335% | 101,956 | 75454 | 0,25993% |
| g-root | 91,547 | 72,573 | 0.20726% | 103,872 | 76349 | 0,26497% |
| h-root | 91,825 | 73,016 | 0.20483% | 102,948 | 76598 | 0,25595% |
| i-root | 91,942 | 73,232 | 0.20349% | 104,017 | 77063 | 0,25913% |
| iad | 92,175 | 73,285 | 0.20493% | 102,324 | 75317 | 0,26393% |
| ihug | 94,719 | 74,549 | 0.21294% | 107,979 | 77792 | 0,27956% |
| k-peer | 91,851 | 72,954 | 0.20573% | 103,672 | 76607 | 0,26106% |
| k-root | 91,726 | 73,100 | 0.20306% | 101,974 | 76451 | 0,25028% |
| lhr | 92,079 | 73,329 | 0.20362% | 101,188 | 75229 | 0,25654% |
| m-root | 92,347 | 73,112 | 0.20829% | 101,321 | 74611 | 0,26361% |
| mwest | 91,525 | 72,722 | 0.20544% | 103,074 | 76388 | 0,25890% |
| nrt | 92,021 | 73,137 | 0.20521% | 101,286 | 74612 | 0,26335% |
| riesling | 90,913 | 72,126 | 0.20664% | 100,426 | 73859 | 0,26454% |
| sjc | 91,459 | 72,762 | 0.20443% | 101,665 | 74936 | 0,26291% |
| uoregon | 90,585 | 72,082 | 0.20426% | 100,851 | 75256 | 0,2537% |
| yto | 91,200 | 72,471 | 0.20536% | 102,625 | 76155 | 0,25792% |

Table 6: Interfaces missed by the ordinary backwards algorithm

# C Searching

## C.1 Redundancy Reduction

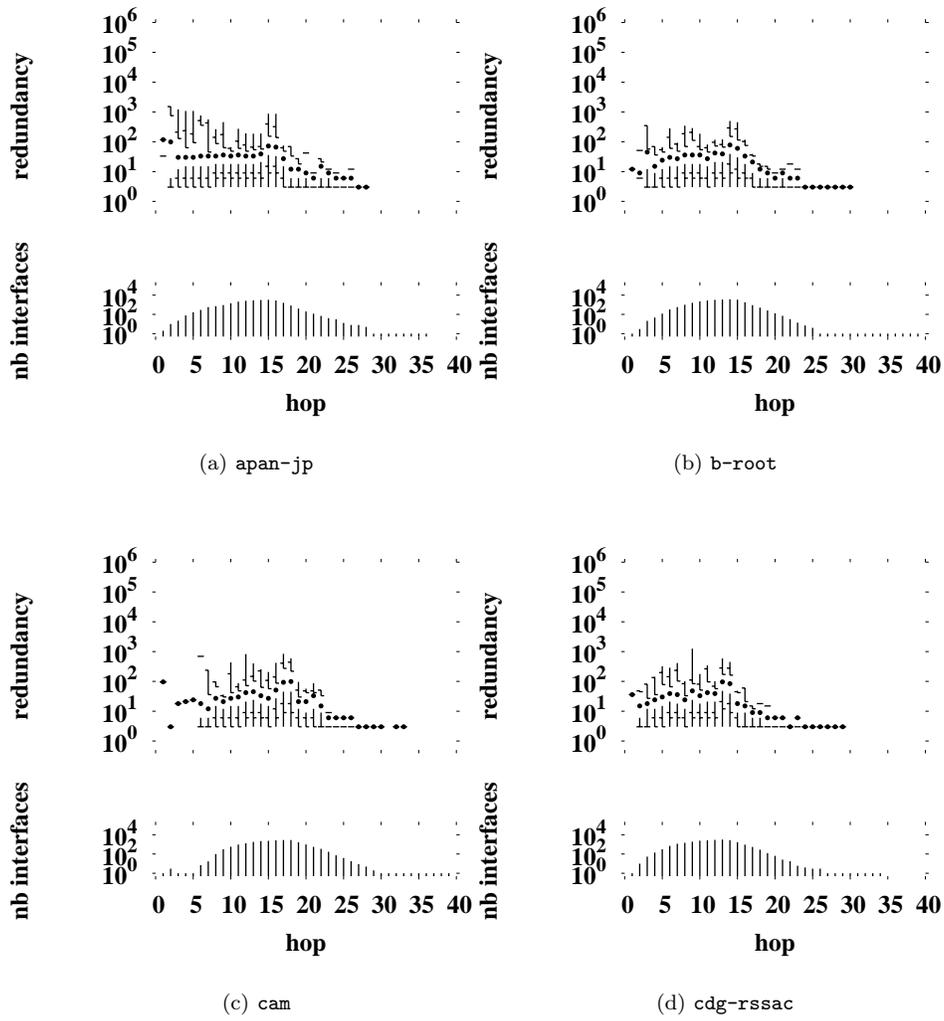

Figure 19: Redundancy when probing with the searching algorithm - 1

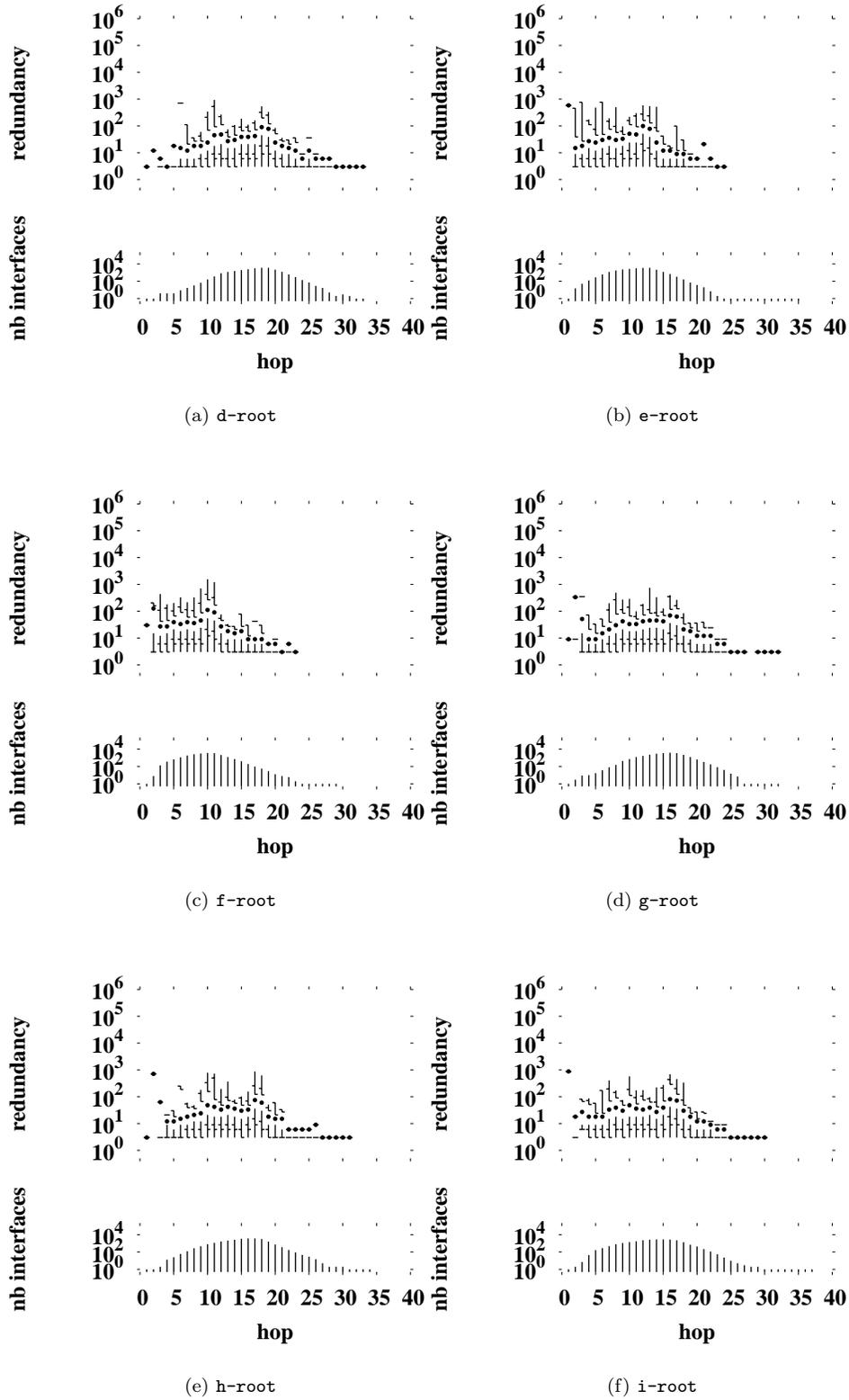

Figure 20: Redundancy when probing with the searching algorithm - 2

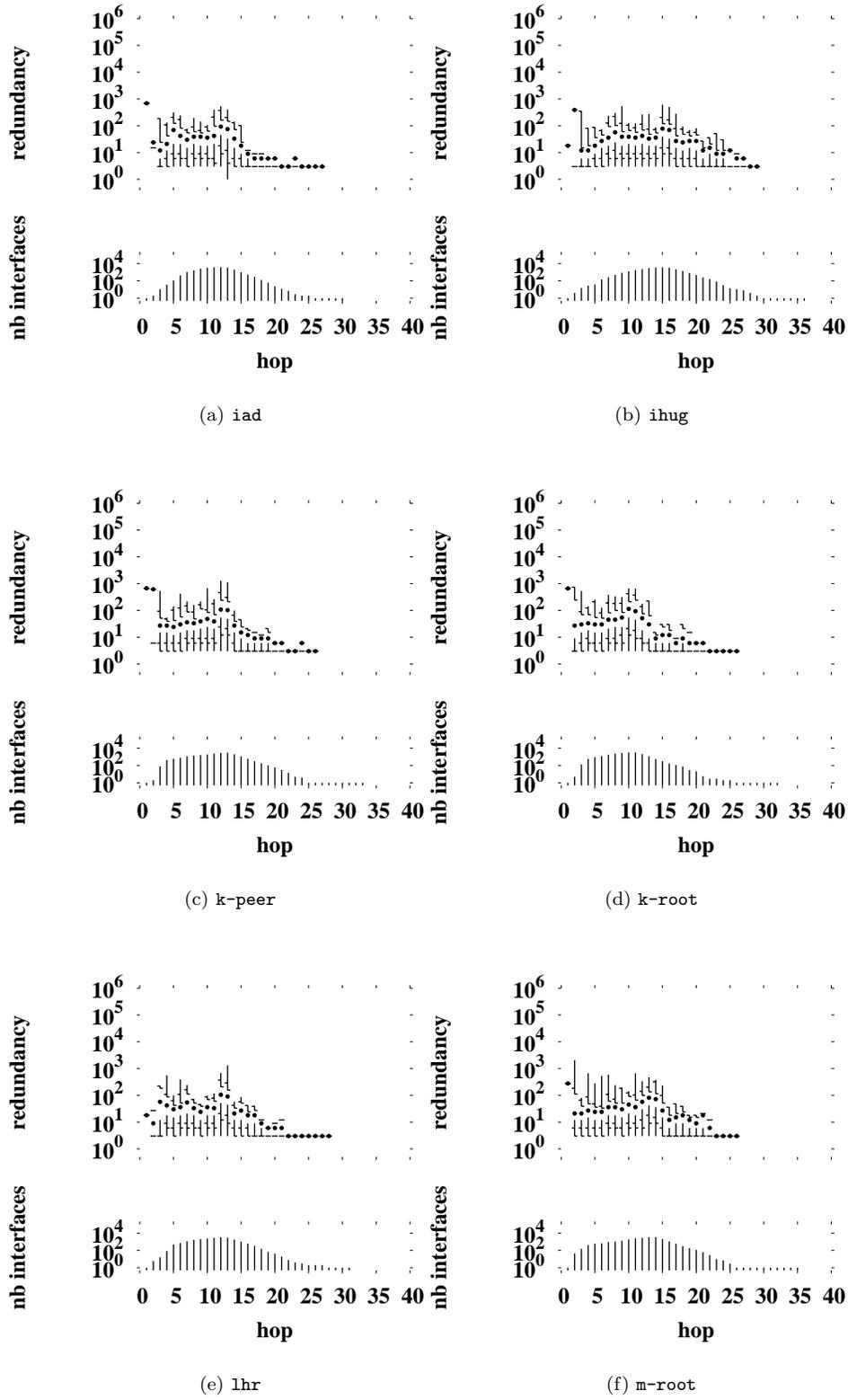

Figure 21: Redundancy when probing with the searching algorithm - 3

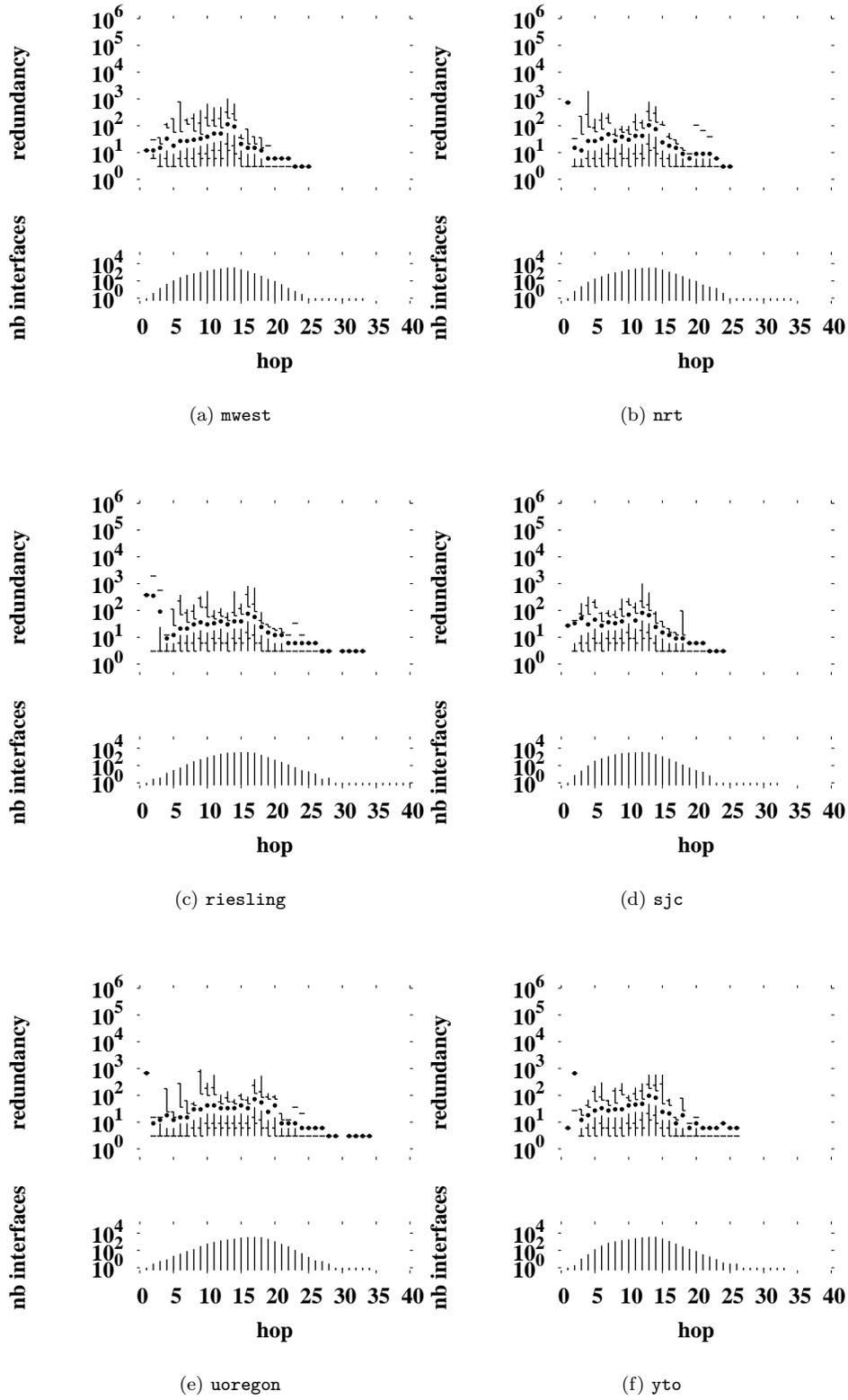

Figure 22: Redundancy when probing with the searching algorithm - 4

## C.2 Incomplete Paths

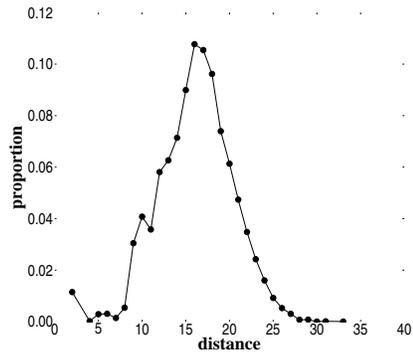
(a) `apan-jp`

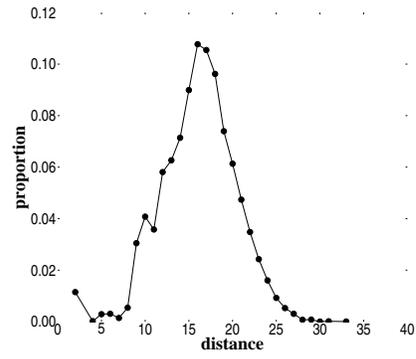
(b) `b-root`

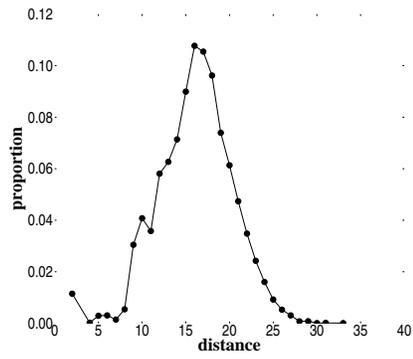
(c) `cam`

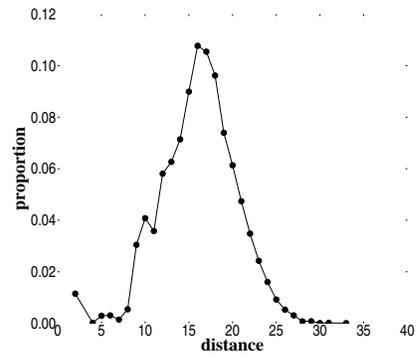
(d) `cdg-rssac`

Figure 23: Incomplete paths distribution - 1

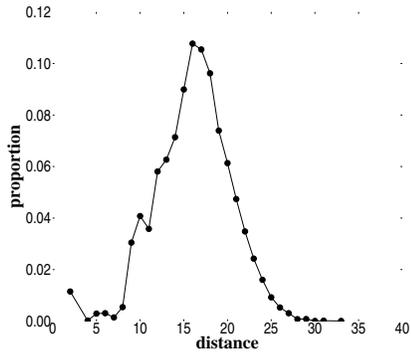
(a) `d-root`

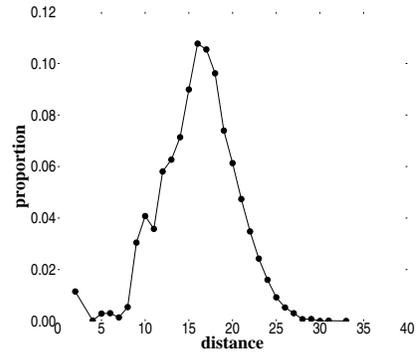
(b) `e-root`

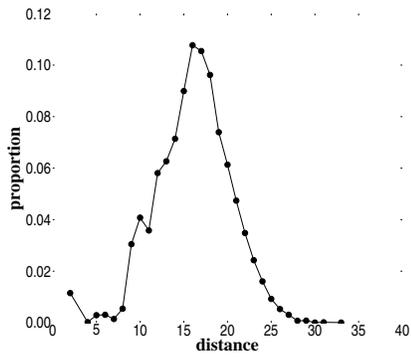
(c) `f-root`

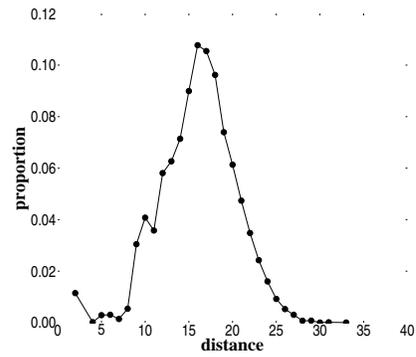
(d) `g-root`

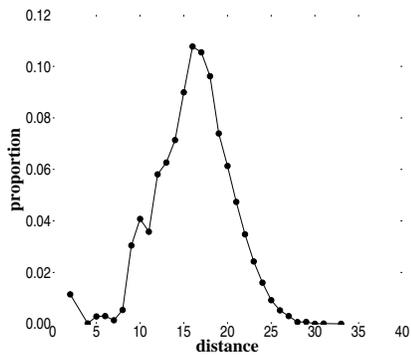
(e) `h-root`

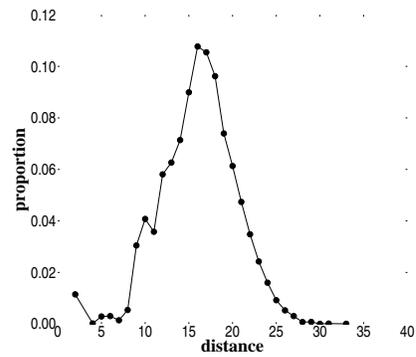
(f) `i-root`

Figure 24: Incomplete paths distribution - 2

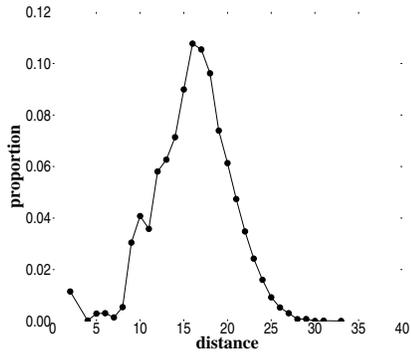
(a) `iad`

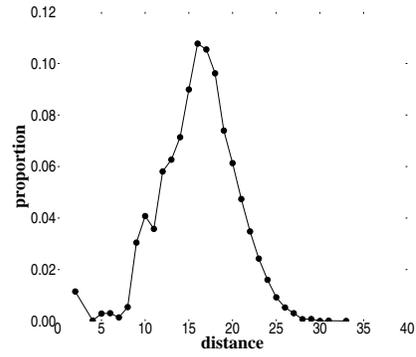
(b) `ihug`

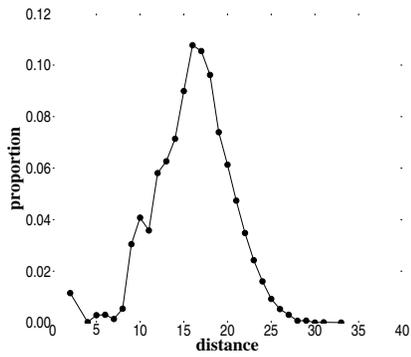
(c) `k-peer`

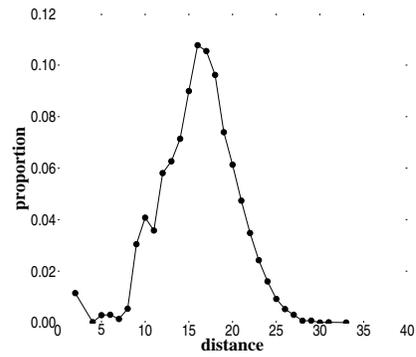
(d) `k-root`

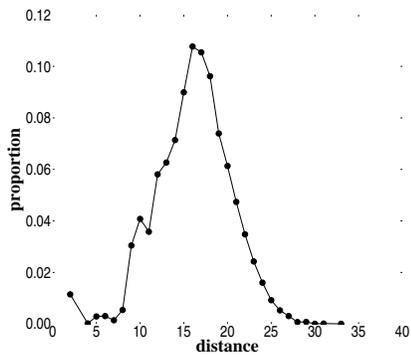
(e) `lhr`

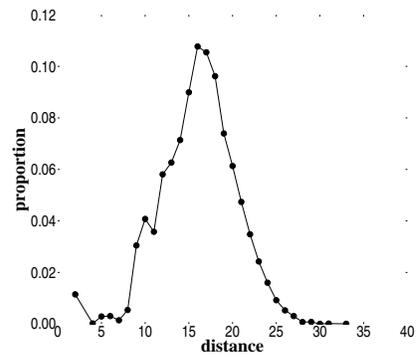
(f) `m-root`

Figure 25: Incomplete paths distribution - 3

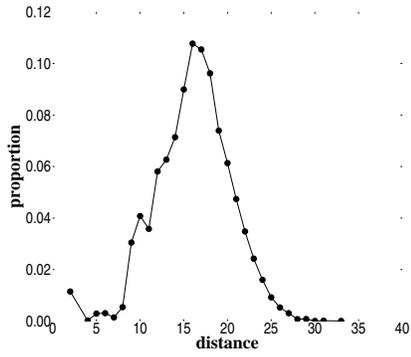
(a) `mwest`

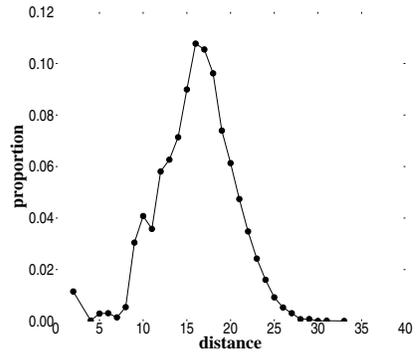
(b) `nrt`

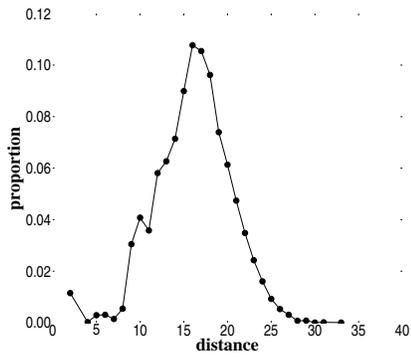
(c) `riesling`

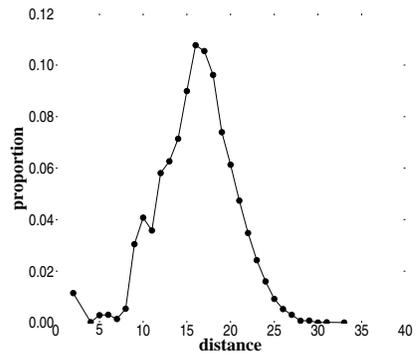
(d) `sjc`

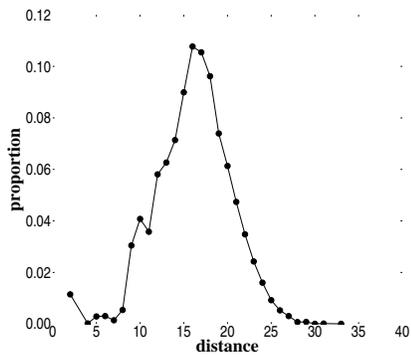
(e) `uoregon`

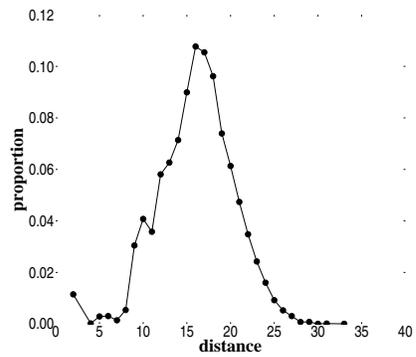
(f) `yto`

Figure 26: Incomplete paths distribution - 4

# D Searching Ordinary Backwards

## D.1 Redundancy Reduction

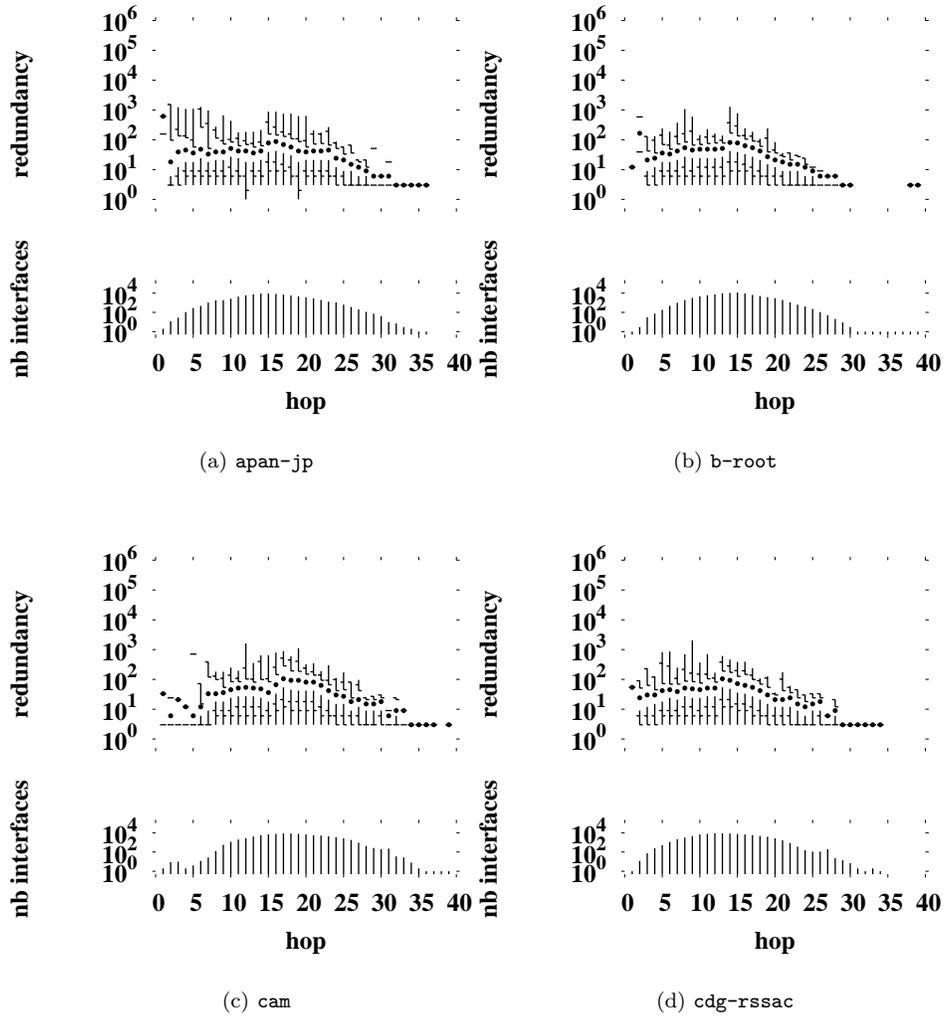

Figure 27: Redundancy when probing with the SearchingOB ordinary backwards algorithm - 1

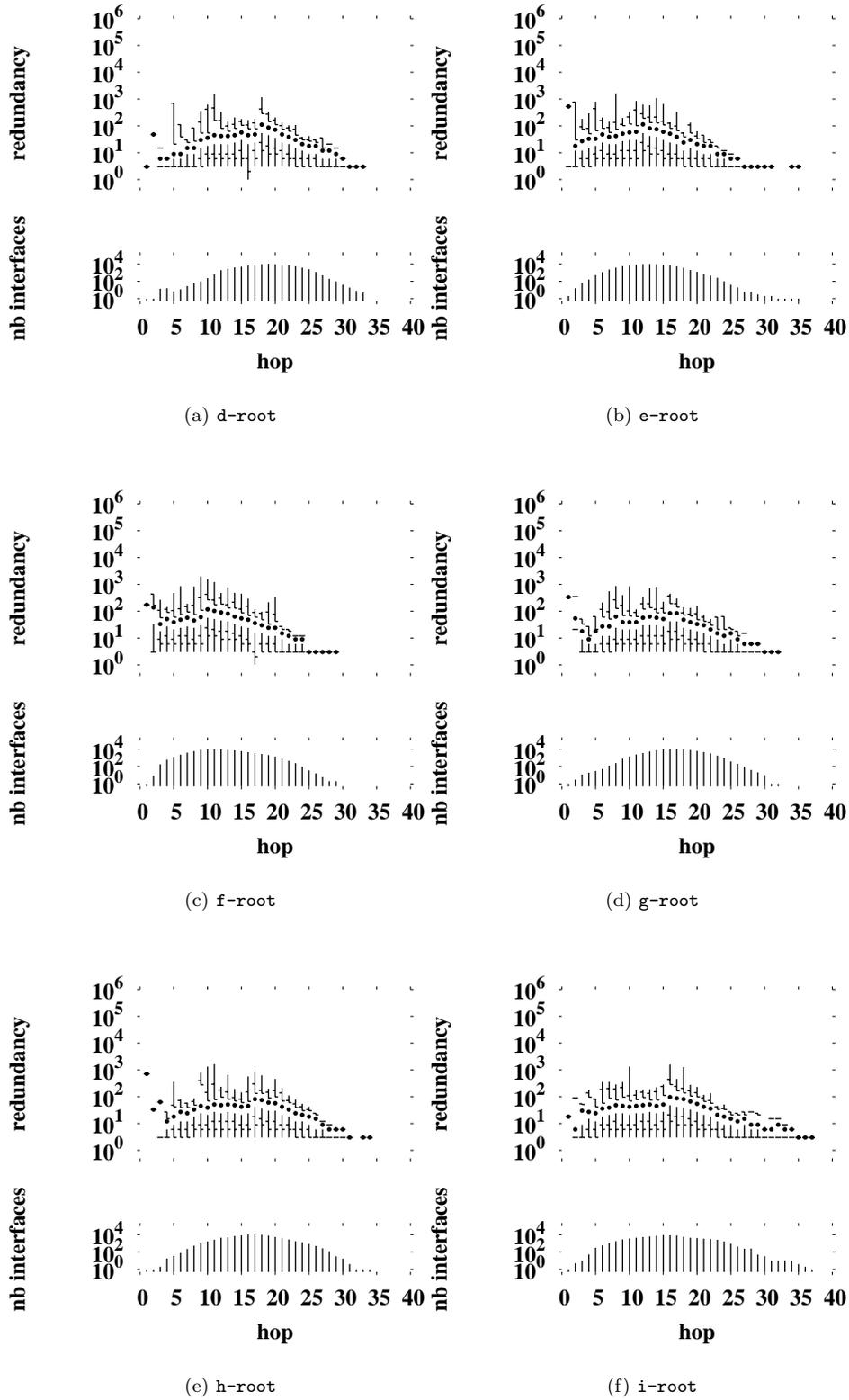

Figure 28: Redundancy when probing with the SearchingOB ordinary backwards algorithm - 2

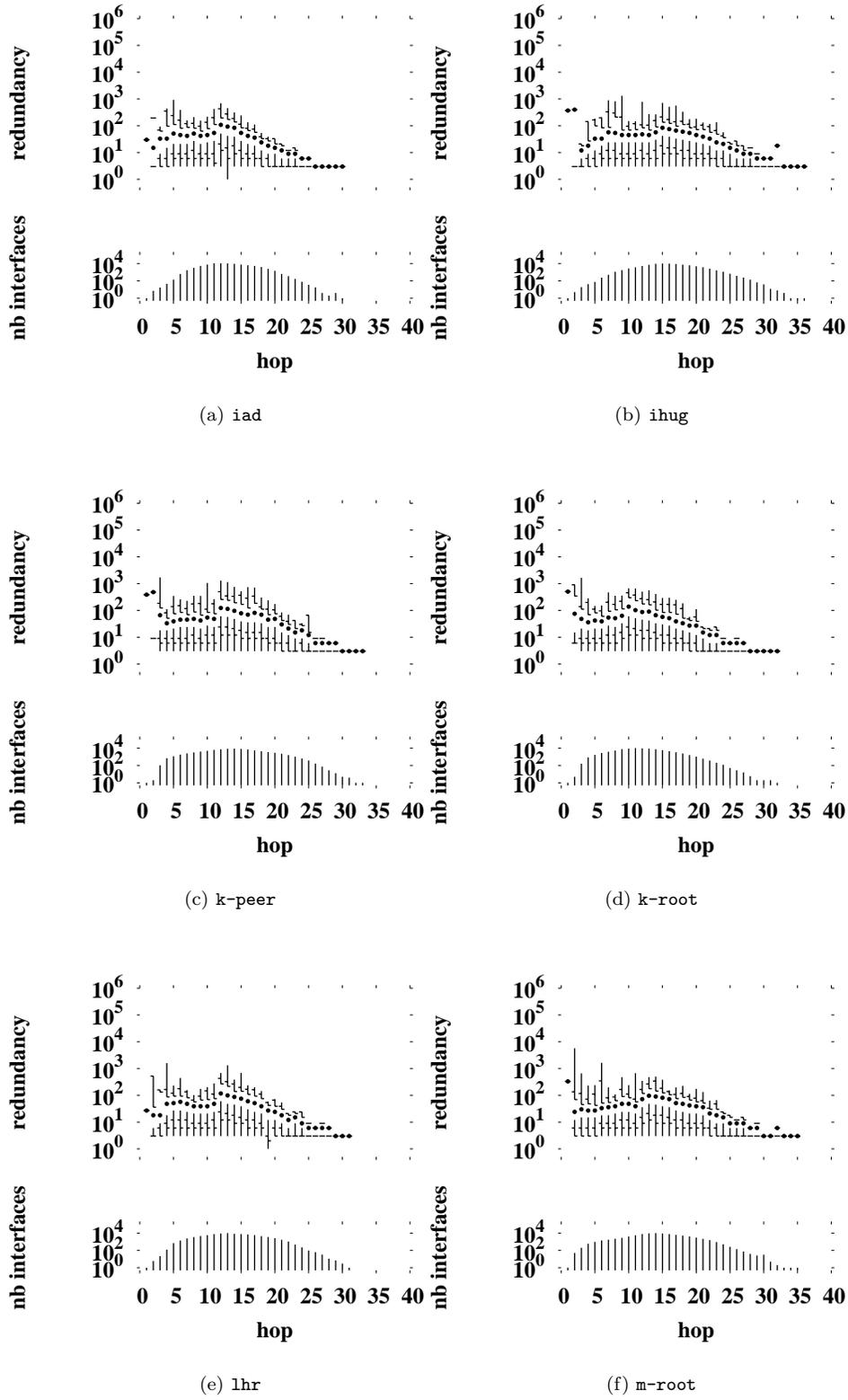

Figure 29: Redundancy when probing with the SearchingOB ordinary backwards algorithm - 3

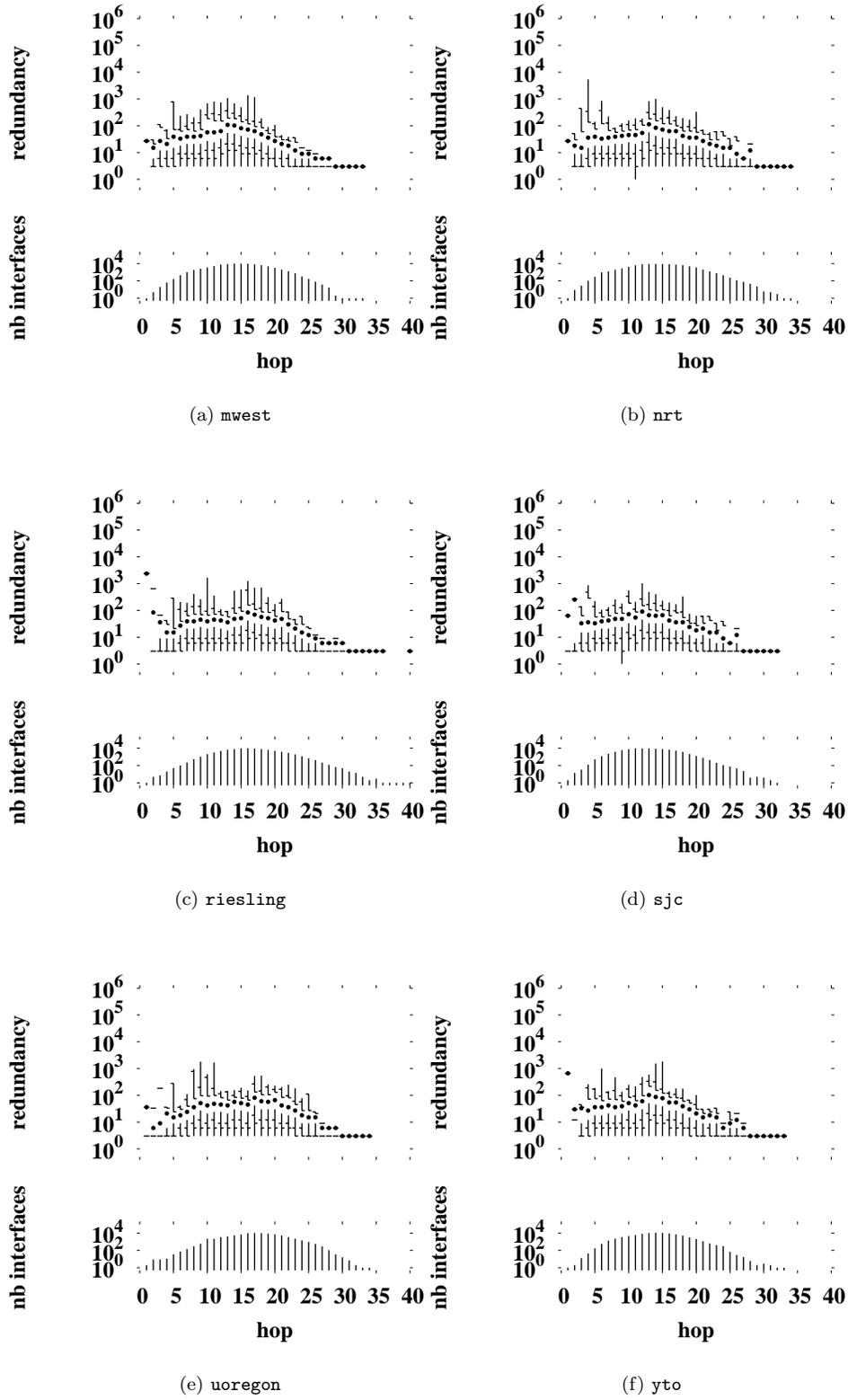

Figure 30: Redundancy when probing with the SearchingOB ordinary backwards algorithm

## D.2 Losses

| Monitor | Interfaces | | | Links | | |
|---|---|---|---|---|---|---|
| | total | discovered | % missed | total | discovered | % missed |
| apan-jp | 86763 | 83,674 | 0.03560% | 96,908 | 88,147 | 0.09040% |
| b-root | 92754 | 89,540 | 0.03465% | 103,595 | 94,469 | 0.08809% |
| cam | 90796 | 88,278 | 0.02773% | 101,068 | 93,822 | 0.07169% |
| cdg-rssac | 90962 | 87,958 | 0.03302% | 100,258 | 92,844 | 0.07394% |
| d-root | 91136 | 89,167 | 0.02160% | 100,821 | 94,634 | 0.06136% |
| e-root | 90952 | 88,831 | 0.02332% | 102,749 | 95,018 | 0.07524% |
| f-root | 92123 | 90,926 | 0.01299% | 101,956 | 95,770 | 0.06067% |
| g-root | 91547 | 89,480 | 0.02257% | 103,872 | 95,335 | 0.08218% |
| h-root | 91825 | 88,914 | 0.03170% | 102,948 | 94,040 | 0.08652% |
| i-root | 91942 | 88,158 | 0.04115% | 104,017 | 94,484 | 0.09164% |
| iad | 92175 | 90,374 | 0.01953% | 102,324 | 95,031 | 0.07127% |
| ihug | 94719 | 91,882 | 0.02995% | 107,979 | 98,129 | 0.09122% |
| k-peer | 91851 | 89,431 | 0.02634% | 103,672 | 96,780 | 0.06647% |
| k-root | 91726 | 89,715 | 0.02192% | 101,974 | 95,884 | 0.05972% |
| lhr | 92079 | 89,877 | 0.02391% | 101,188 | 94,754 | 0.06358% |
| m-root | 92347 | 89,448 | 0.03139% | 101,321 | 93,277 | 0.07939% |
| mwest | 91525 | 89,953 | 0.01717% | 103,074 | 96,561 | 0.06318% |
| nrt | 92021 | 89,825 | 0.02386% | 101,286 | 93,955 | 0.07237% |
| riesling | 90913 | 87,681 | 0.03555% | 100,426 | 91,518 | 0.08870% |
| sjc | 91459 | 89,299 | 0.02361% | 101,665 | 93,713 | 0.07821% |
| uoregon | 90585 | 87,990 | 0.02864% | 100,851 | 92,971 | 0.07813% |
| yto | 91200 | 89,541 | 0.01819% | 102,625 | 95,834 | 0.066173% |

Table 7: Interfaces missed by the SearchingOB ordinary backwards algorithm

# E    Algorithms Comparison

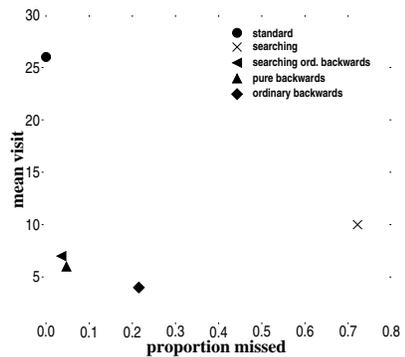
(a) `apan-jp`

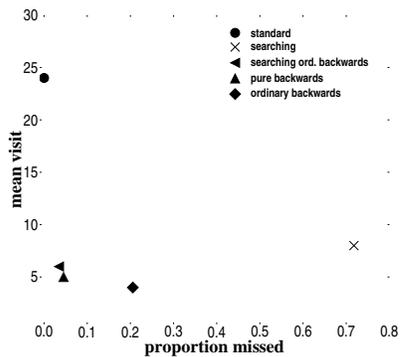
(b) `b-root`

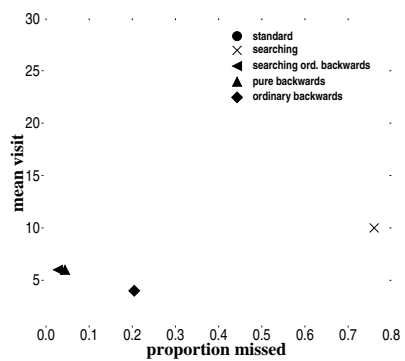
(c) `cam`

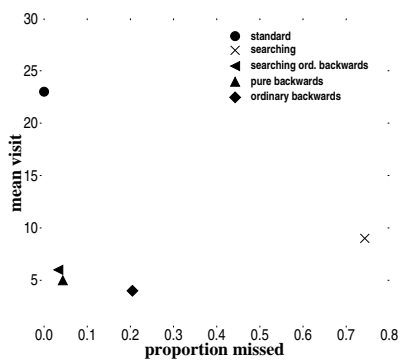
(d) `cdg-rssac`

Figure 31: Algorithms comparison - 1

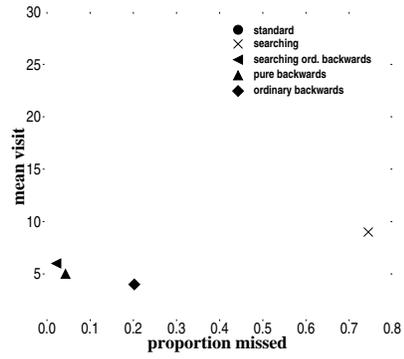
(a) `d-root`

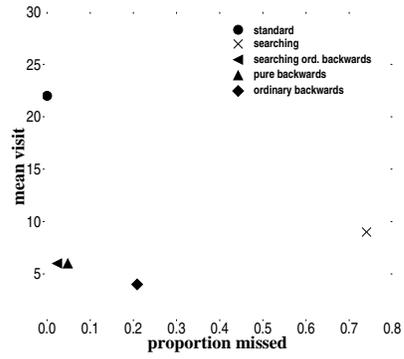
(b) `e-root`

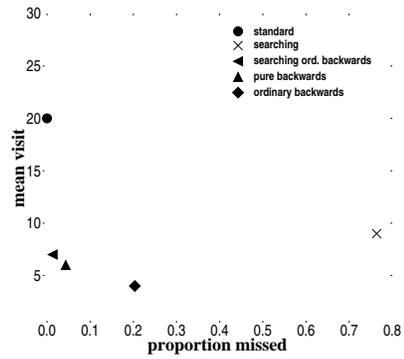
(c) `f-root`

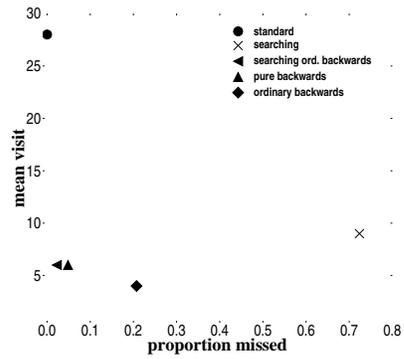
(d) `g-root`

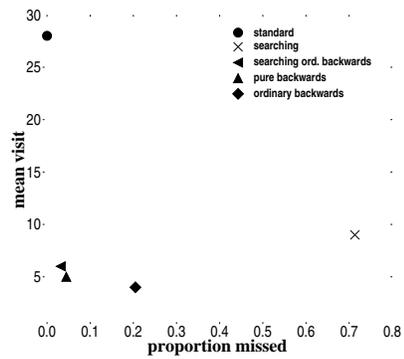
(e) `h-root`

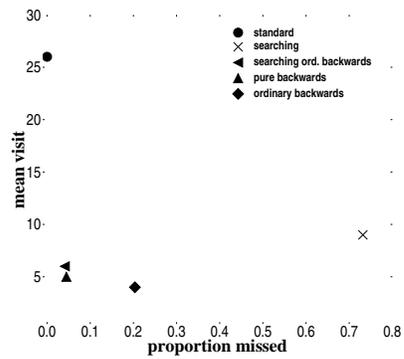
(f) `i-root`

Figure 32: Algorithms comparison - 2

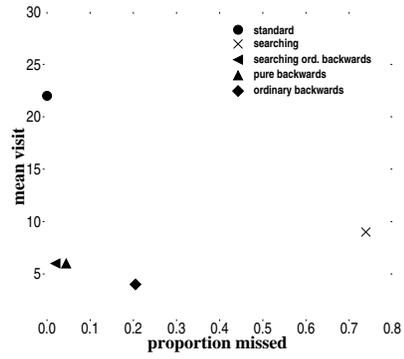

(a) `iad`

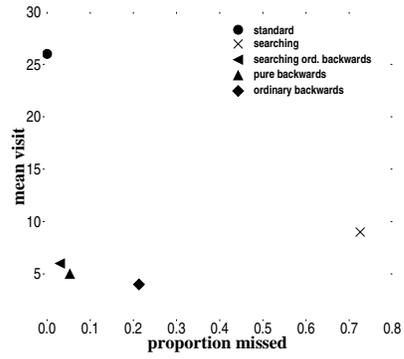

(b) `ihug`

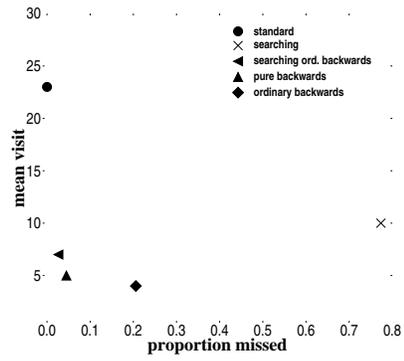

(c) `k-peer`

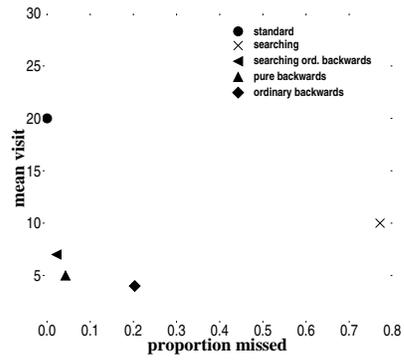

(d) `k-root`

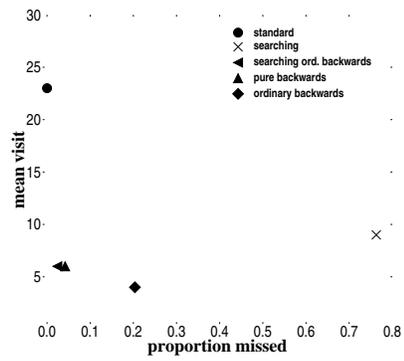

(e) `lhr`

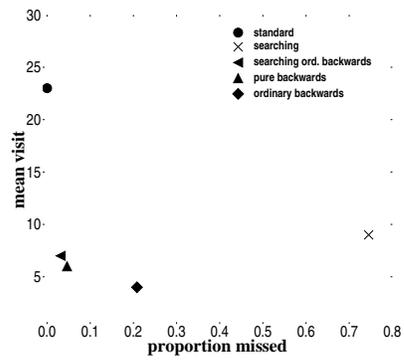

(f) `m-root`

Figure 33: Algorithms comparison - 3

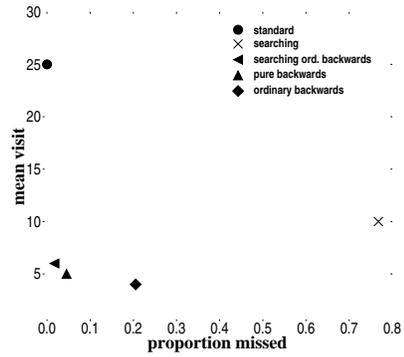

(a) `mwest`

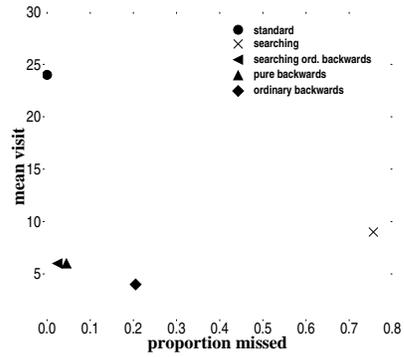

(b) `nrt`

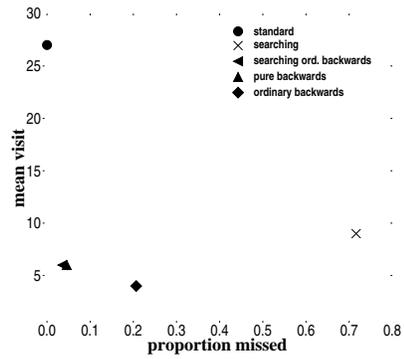

(c) `riesling`

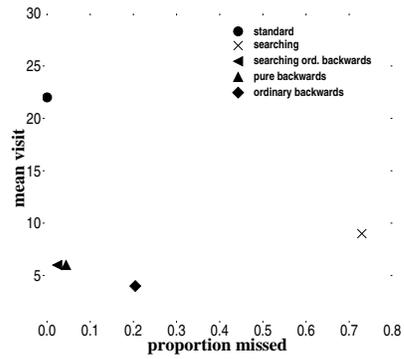

(d) `sjc`

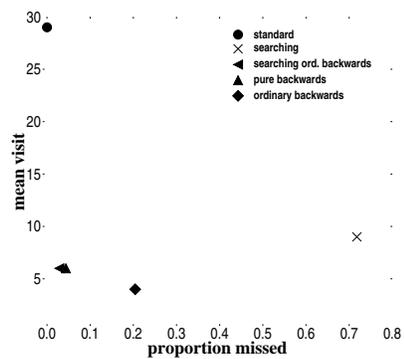

(e) `uoregon`

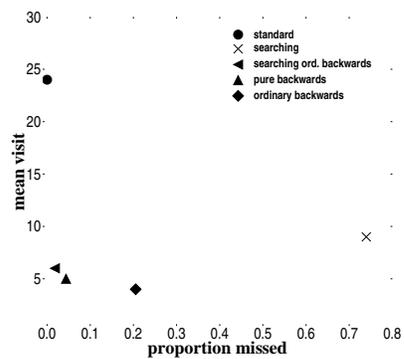

(f) `yto`

Figure 34: Algorithms comparison - 4